\documentclass[twocolumn]{aastex62}
\usepackage{l3regex}
\usepackage[version=3]{mhchem}
\usepackage{subfigure}

\newcommand{\de}{\,\mathrm{d}}

\graphicspath{{./}{figures/}}

\accepted{to ApJ on June 20, 2019}

\shorttitle{Modeling C-Shock Chemistry in Isolated Molecular Outflows}
\shortauthors{Burkhardt et al.}

\begin{document}

\title{Modeling C-Shock Chemistry in Isolated Molecular Outflows}

\correspondingauthor{Andrew M. Burkhardt}
\email{andrew.burkhardt@cfa.harvard.edu}

\author[0000-0003-0799-0927]{Andrew M. Burkhardt}
\altaffiliation{Submillimeter Array (SMA) Postdoctoral Fellow}
\affiliation{Harvard-Smithsonian Center for Astrophysics, Cambridge, MA 02138, USA}
\affiliation{Astronomy Department, University of Virginia, Charlottesville, VA 22904, USA}

\author[0000-0002-5171-7568]{Christopher N. Shingledecker}
\affiliation{Max-Planck-Institut f\"{u}r extraterrestrische Physik, D-85748 Garching, Germany}
\affiliation{Institute for Theoretical Chemistry, University of Stuttgart, Pfaffenwaldring 55, 70569, Germany}
\affiliation{Chemistry Department, University of Virginia, Charlottesville, VA 22904, USA}

\author[0000-0003-1837-3772]{Romane Le Gal}
\affiliation{Harvard-Smithsonian Center for Astrophysics, Cambridge, MA 02138, USA}

\author[0000-0003-1254-4817]{Brett A. McGuire}
\altaffiliation{Hubble Fellow of the National Radio Astronomy Observatory}
\affiliation{Harvard-Smithsonian Center for Astrophysics, Cambridge, MA 02138, USA}
\affiliation{National Radio Astronomy Observatory, Charlottesville, VA 22904, USA}

\author[0000-0001-9479-9287]{Anthony J. Remijan}
\affiliation{National Radio Astronomy Observatory, Charlottesville, VA 22904, USA}

\author[0000-0002-4649-2536]{Eric Herbst}
\affiliation{Astronomy Department, University of Virginia, Charlottesville, VA 22904, USA}
\affiliation{Chemistry Department, University of Virginia, Charlottesville, VA 22904, USA}



\begin{abstract}
Shocks are a crucial probe for understanding the ongoing chemistry within ices on interstellar dust grains where many complex organic molecules (COMs) are believed to be formed. However, previous work has been limited to the initial liberation into the gas phase through non-thermal desorption processes such as sputtering. Here, we present results from the adapted three-phase gas-grain chemical network code \textsc{nautilus}, with the inclusion of additional high-temperature reactions, non-thermal desorption, collisional dust heating, and shock-physics parameters. This enhanced model is capable of reproducing many of the molecular distributions and abundance ratios seen in our prior observations of the prototypical shocked-outflow L1157. In addition, we find that, among others, NH$_2$CHO, HCOOCH$_3$, and CH$_3$CHO have significant post-shock chemistry formation routes that differ from those of many other COMs observed in shocks. Finally, a number of selected species and phenomena are studied here with respect to their usefulness as  shock tracers in various astrophysical sources.

\end{abstract}



\section{Introduction} \label{sec:intro}
With the advent of modern observatories such as the Atacama Large Millimeter/ submillimeter Array (ALMA), there has been an explosion in high-resolution, high-sensitivity observations of star and planet-forming regions, with an emphasis on the role of chemical inventories on this evolution \citep{Belloche:2016fm,Jorgensen:2016cq}. This is highlighted by a surge of interest in the interaction of gas and ice chemistry, including the condensation fronts (snow lines) within protoplanetary disks \citep{Qi:2013dx,walsh_complex_2014,bergner_methanol_2017} and the thermal warm-up in hot cores and corinos \citep{Belloche:2016fm,Jorgensen:2016cq}. In particular, it is believed many complex organic molecules (COMs) are efficiently formed primarily in interstellar ices \citep{Garrod:2013id,Herbst:2009go}. However, because these frozen molecules cannot be observed through radio (rotational) spectroscopy, our knowledge of interstellar ice chemistry remains highly unconstrained. These difficulties in deducing the molecular composition of ices directly means that we must rely on, to date, unconstrained astrochemical models to make inferences. Refinement of these models is therefore critical to unlock the potential of modern astrochemical observations. One way this can be accomplished is to indirectly observe ice abundances in regions where the ice is actively being liberated into the gas phase, and compare these abundances to the models using those observations. In these types of sources, it is important that additional chemical processing does not dominate the desorption process. Low-velocity shocks, which liberate molecules off of the ice and into the gas phase (e.g. sputtering) without destroying most molecular bonds, provide this indirect probe \citep{RequenaTorres:2006ki}. Because shocks are ubiquitous, but transient, features throughout the interstellar medium (ISM), it is crucial to understand the physico-chemical evolution of these phenomena.\\

One of the most promising locations to study shocks is the low-mass protostar L1157, an isolated, nearby ($\sim$250\,pc\,\citep{Looney:2007}) source with a prominent bipolar outflow where the ejection events have shocked the surrounding dark cloud, including the pristine, chemically rich ice \citep{Arce:2008ha,Benedettini:2013,Burkhardt:2016bs,Lefloch:2017}. Here, we can study the effects that shocks have on a material in relative isolation of other competing processing (e.g. thermal processing common in hot cores). Prior observations by \citet{Burkhardt:2016bs} of the bow shock fronts in this source have shown that the chemical inventory affected by the shock could be divided into three classes: those directly sputtered from grains (e.g. \ce{CH3OH}), those already in the gas but enhanced by the shock (e.g. \ce{HCO^+}), and those that were sputtered and subsequently enhanced by gas-phase chemistry (e.g. HNCO). To date, only a small number of these species and a few more complex molecules have been targeted by interferometric observations in shocks (see, e.g., \citep{Benedettini:2013,Codella:2017ij,Kwon:2015}). However, single-dish surveys have revealed that the bow shocks within this molecular outflow are extremely chemically rich, with the detection of complex molecules including NH$_2$CHO, CH$_3$CHO, CH$_3$OCH$_3$, HCOOCH$_3$, C$_2$H$_5$OH, and CH$_3$CN \citep{Arce:2008ha,Lefloch:2017}. \\

In this paper, we present an adaptation of the gas-grain chemical network code \textsc{nautilus} \citep{Ruaud:2016bv} in order to study how complex chemistry is impacted by C-shocks. The overall description of the model is discussed in Section \ref{sec:model}, the results from these models are presented in Section \ref{sec:p5_results}, and a discussion follows in Section \ref{sec:discussion}.

\section{Model} \label{sec:model}


In order to study the competitive effects of sputtering and post-shock chemistry in these types of sources, we have adapted the three-phase (i.e. gas, ice-surface, and ice-bulk) gas-grain chemical network code \textsc{nautilus} (version 1.1) \citep{Ruaud:2016bv}, which calculates the molecular abundances as a function of time through a rate-equation approach. The physical parameters of the model were computed to accurately simulate the conditions within astrophysical C-shocks, as discussed in Section \ref{sec:modelstructure}. The measured and computed kinetic rate coefficients for the chemical reactions used are based on the \textsc{KIDA-2014} network \citep{wakelam_2014_2015}, to which the reactions discussed in Section \ref{sec:hitemprx} have been added.

\subsection{Structure of Model}\label{sec:modelstructure}

\begin{figure*}
\includegraphics[width=\textwidth]{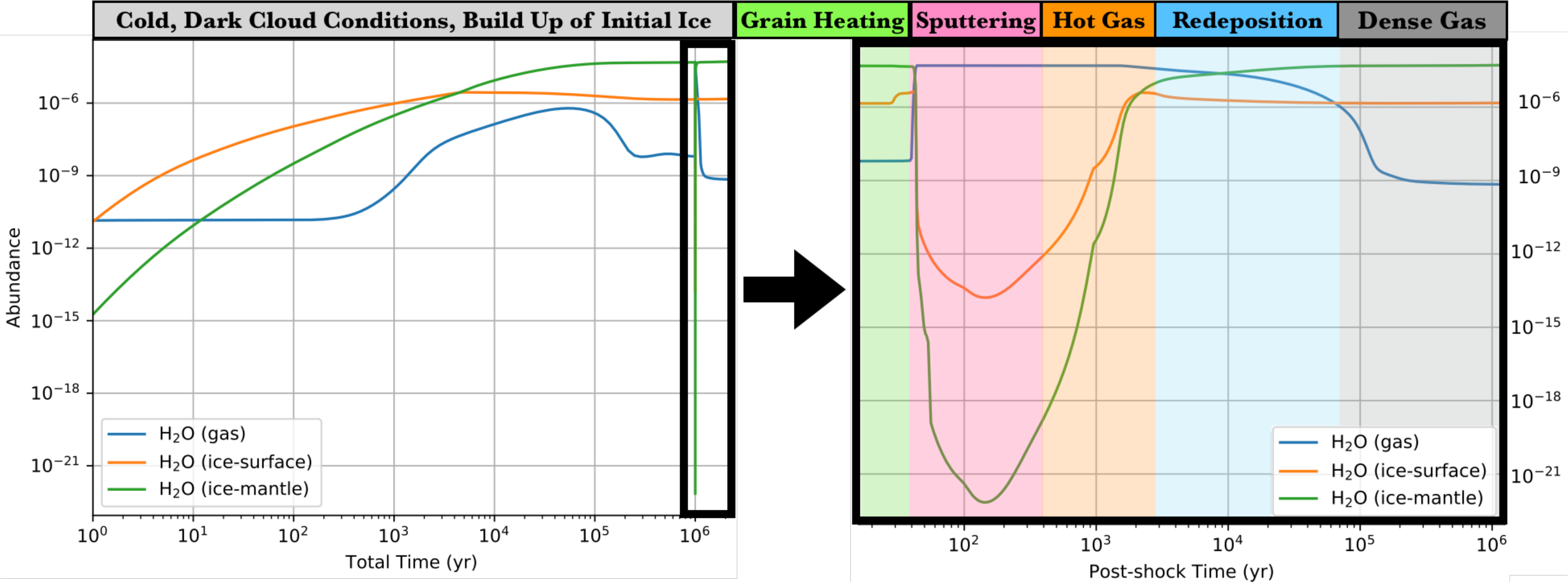} 
\caption{\textit{Left} The temporal three-phase abundances of H$_2$O relative to hydrogen with respect to total time within the model, where the ice and gas abundances are built up to resemble a typical dark cloud. \textit{Right} The temporal three-phase abundances of H$_2$O relative to hydrogen with respect to the time following the shock. Here, $t$=0 corresponds to 10$^6$ years after the start of the left figure, with a rescaling of the logarithm in the abscissa, zooming in on the shock event. On the top, the general time regimes of differing physical processing are labeled and colored.  }
\label{fig:modeloutline}
\end{figure*}

As shown in Figure \ref{fig:modeloutline}, the first stage of the model was initially run for 10$^6$ years using physical conditions appropriate for cold cores such as TMC-1 \citep{Hincelin:2011fr,Loomis:2016jsa,mcguire_detection_2017,mcguire_detection_2018}. Specifically, the constant physical conditions over this time scale are:
\begin{align}
	T_{\text{gas},0} = T_{\text{dust},0} &= 10 \text{ K}\\
    n_{\text{H}_2,0} &= 5\times10^4 \text{ cm}^{-3} \\
    A_{V,0} &= 10 \\
    \zeta_{\text{CR}} & = 1.3\times10^{-17} \text{ s}^{-1}
\end{align}
The initial elemental abundances are given in Table \ref{tab:elemental_ab}. During this initial phase, the grains are able to build up a sufficiently thick ice mantle that is representative of a cold core prior to a shock event, and the majority of the gas and solid-phase abundances have relatively stable values. Due to the difficulties of astronomical observations of interstellar ices, the measured solid-phase abundances in the ISM are highly unconstrained and limited to only the most abundant species with identified infrared absorption features \citep{Boogert:2015fx}. As such, the subsequent analysis of abundances following the liberation from the ice will be biased by this large source of uncertainties. If one assumes that the vast majority of the solid-phase abundance is lifted into the gas-phase and that the abundances of these species experience little gas-phase modification during the shock, then the initial ice abundance will be a 1:1 tracer of the post-shock gas-phase abundance enhancement. However, it is likely that many species will additionally undergo some form of chemistry during or after the shock. It is also possible that a non-trivial fraction of the sputtered species may dissociate during liberation \citep{Suutarinen:2014}. Thus, better constraining the effects of these competing processes will allow us to disentangle the sources of molecular enhancements or diminishment. Due to the previously noted uncertainties in the ice chemistry, though, in this work we will generally focus on the overall temporal evolution of the abundances and the regimes in which significant enhancements occur.

\begin{table}
\centering
\scriptsize
\caption{Initial Elemental Abundances}
\begin{tabular}{@{\hskip3pt}c@{\hskip3pt} c c c @{\hskip1pt}c@{\hskip1pt}}
\hline\hline
Species & $n_i/n_{\text{H}}$    & Reference\\
\hline
H$_2$   & 0.5\\
He      & 9.0$\times$10$^{-2}$              & 1\\
C$^+$   & 1.7$\times$10$^{-4}$              & 2 \\
N       & 1.7$\times$10$^{-5}$              & 2\\ 
O       & 2.4$\times$10$^{-4}$              & 3\\ 
S$^+$   & 8.0$\times$10$^{-8}$              & 4 \\
Si$^+$  & 8.0$\times$10$^{-9}$              & 4 \\
Fe$^+$  & 3.0$\times$10$^{-9}$              & 4 \\
Na$^+$  & 2.0$\times$10$^{-9}$              & 4 \\
Mg$^+$  & 7.0$\times$10$^{-9}$              & 4 \\
P$^+$   & 2.0$\times$10$^{-10}$             & 4 \\
Cl$^+$  & 1.0$\times$10$^{-9}$              & 4 \\
F$^+$   & 6.68$\times$10$^{-9}$             & 5 \\
\multicolumn{3}{l}{[1] \citet{wakelam_polycyclic_2008}}\\
\multicolumn{3}{l}{[2] \citet{jenkins_unified_2009}}\\
\multicolumn{3}{l}{[3] \citet{Hincelin:2011fr}}\\
\multicolumn{3}{l}{[4] \citet{graedel_kinetic_1982}}\\
\multicolumn{3}{l}{[5] \citet{Neufeld:2005lk} }
\label{tab:elemental_ab}
\end{tabular}
\end{table}

Following the initial build up of ice for 10$^6$ years, we introduce a shock event into the model, as described in Section \ref{sec:shockstructure}. The evolution of the shock event takes roughly 10$^5$ years, depending on the velocity of the shock and initial conditions of the cloud.  Here, in lieu of direct abundance comparisons, we focus on the general enhancements of the species from the pre-shock abundances to the post-shock abundances. Because the sputtering occurs much faster than the post-shock chemistry, it is possible for some species to easily differentiate which process significantly dominates the enhancements. 

After material cools back down to the dark cloud conditions, we then continue to run the model until approximately 10$^6$ years after the shock event. This facilitates the study of the after-effects on the chemistry, long after the shock has passed. Any significant trends seen here may point to important signatures that would indicate if the material was recently shocked, compared to what was seen in the pre-shock conditions. This will be discussed in detail in Section \ref{sec:p5_results}.

\subsection{Adaptation of \textsc{nautilus} for Shock Chemistry}
To accurately simulate the chemistry within shocked media, the \textsc{nautilus} code and KIDA-2014 network needed to be enhanced to include a number of additional shock-related processes and reactions. These include additional high-temperature reactions from \citet{Harada:2010} and \citet{Garrod:2013id}, the physical structure and sputtering processes as described by \citet{jimenezserra:2008}, and the evolution of the dust temperature as described by \citet{Miura:2017}, of which all will be described in detail in the subsequent sections. 
The high degree of chemical complexity within the network used, especially for the ice chemistry, allows us to simulate the chemistry of much larger COMs than previous models \citep{Lesaffre:2013,2017A&A...603A.105D,Codella:2017ij}.  

\subsubsection{Physical Structure within C-Shocks} \label{sec:shockstructure}
As extensively studied previously, the physical conditions within shocks and turbulent sources are found to have a significant impact on the subsequent chemistry \citep{Gusdorf:2015dta,Lefloch:2017,Codella:2017ij,Burkhardt:2016bs}. While several studies looked into these effects by coupling a chemical network to magnetohydrodynamics (MHD) simulations \citep{hincelin_survival_2013}, the chemical networks used were often limited in scope in order to maintain a reasonable computation time. In order to leverage the advanced gas-grain chemical network of \textsc{osu.kida} within \textsc{nautilus}, the physical conditions are simplified to a single, 1-D slab. To do this, we adopted a system of parametric equations developed by \citet{jimenezserra:2008} that have been fit to match the simulated physical conditions by the MHD shock-structure calculated by \citet{Flower:2003} and \citet{Kaufman:1996}. By using these parametric equations, we are able to accurately reproduce these physical conditions while at the same time treating the chemistry in more detail than in previous studies.

Here, we consider a plane-parallel C-shock where accelerated material interacts with a cold, quiescent cloud. Because these equations were fit to data from the numerical simulations, the dependent variable is the spatial location within a snapshot of the shock, $z$. Since the pre-shocked material is assumed to be homogeneous, these physical locations can be translated to corresponding times it would take to propagate through the shock to achieve the spatial coordinate given by assuming a single velocity to describe the aggregate speed of the shock (i.e. the shock velocity, $v_s$). It should be noted that the time described here refers to the post-shock time (i.e. beginning on Figure \ref{fig:modeloutline} \textit{Right}). Across this shock, the magnetic fields within the source will accelerate the charged particles. Because the dust grains tend to be charged in these sources, this causes the predominantly neutral gas to temporarily decouple from the dust within the shock. These velocities of the gas and grains can be described in the frame of reference of the downstream material as
\begin{equation}
	v_{\text{gas/grains}} = (v_s - v_0) \left( 1 - \frac{1}{\cosh\left[ (z-z_0)/z_{\text{gas/grains}} \right]} \right)
\end{equation}
where $v_0$ is the post-shock velocity of both the gas and dust in the co-moving shock frame of reference (i.e. the difference between the peak drift velocity and the overall shock velocity), $v_s$ is the shock velocity, $z_0$ is the distance where the decoupling begins, and $z_{\text{gas}}$ and $z_{\text{grains}}$ are parameters that set the length scale over which neutral gas and grains are decelerated in the shock, respectively. The relative velocity between these two populations, referred to the drift velocity, is then simply
\begin{equation}
	v_d \equiv \lvert v_{\text{gas}} - v_{\text{grains}} \rvert.
\end{equation}
With these velocities, we can covert the spatial coordinate within the shock to the flow of time a gas particle experiences as the shock propagates through it as  
\begin{equation}
	t_{\text{postshock}} = \int \frac{1}{v_s - v_{\text{gas}}} \de z
\end{equation}
From the subsequent time-dependent velocity profile, an initial gas density, $n_{0}$, and initial gas temperature, $T_{\text{gas},0}$, we can develop the temporal evolution of the density and gas temperature. By assuming conservation of mass, one can describe the density as a function of the velocity (and thus time) such that
\begin{equation}
	n_{\text{gas}}(t_{\text{postshock}}) = \frac{v_s}{v_s - v_{\text{gas}}} n_0.
\end{equation}
Since the extinction, $A_V$, can be estimated to be proportional to the gas density \citep{Bergin:1998}, it can be similarly scaled from an initial extinction value $A_{V,0}$.

For the purposes of matching the MHD simulations, the gas temperature is described as a ``Planck-like'' function such that
\begin{equation}
T_{\text{gas}} = T_{\text{gas},0} + \frac{\left(a_T(z - z_0)\right)^{b_T}}{e^{(z-z_0)/z_T}-1}
\end{equation}
where $a_T$ and $z_T$ are fitting parameters to the location and value of the peak gas temperature and $b_T$ is an integer fitting the temperature's sensitivity to shock location. 

For the purposes of this paper, we seek to model the prototypical shocked-outflow L1157. Therefore, while this model is capable of simulating a range of shock velocities and gas densities, we will focus on a detailed analysis of the physical conditions based on derived values from \citet{Burkhardt:2016bs}. And so, we assume a $v_s$ of 20 km s$^{-1}$ and $n_0$ of 5$\times$10$^4$ cm$^{-3}$. The effects of varying these parameters is discussed further in Section \ref{sec:discussion}. Assuming standard dark cloud conditions in the pre-shock gas, our initial gas and dust temperatures were both 10 K \citep{hincelin_new_2015}. From Table 4 from \citet{jimenezserra:2008}, these initial conditions correspond to fitted values of $z_{\text{gas}}=1.4\times$10$^{16}$ cm, $z_{\text{grains}}=3.2\times$10$^{15}$ cm, $z_{\text{T}}=5.0\times$10$^{15}$ cm, and $a_{\text{T}}=2.9\times$10$^{-16}$ K$^{1/6}$ cm$^{-1}$, assuming $z_0=$0 cm.

\begin{figure}
\centering
\includegraphics[width=\columnwidth]{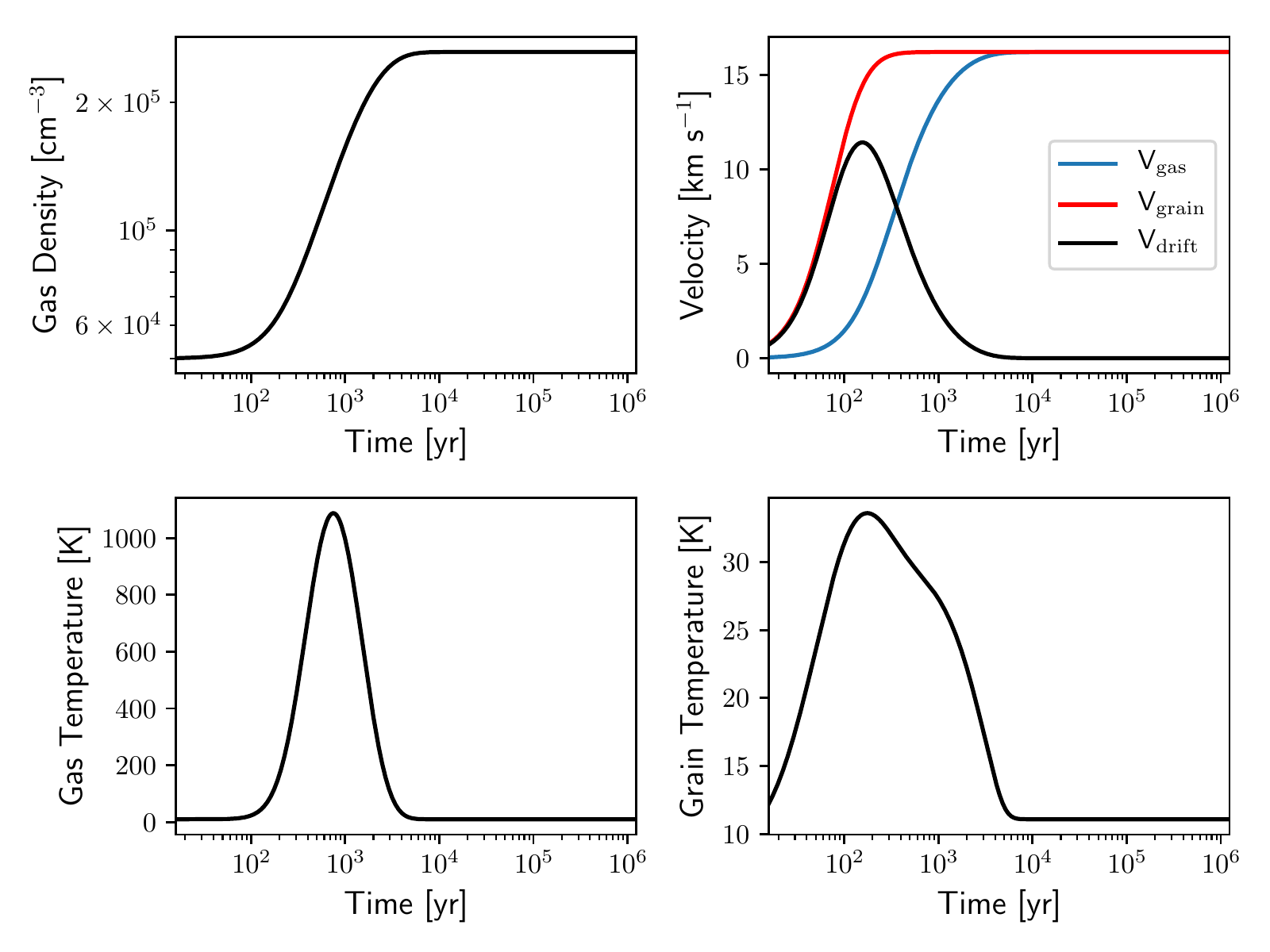} 
\caption{Physical evolution within the shock (i.e. $t_{\text{postshock}}$=0 years corresponds to total time in model of $t_{\text{total}}$=10$^6$ years) for the gas density, transverse velocities, gas temperature, and grain temperature, as described in Section \ref{sec:tdust}. }
\label{fig:shock}
\end{figure}

Using this formalism, the physical conditions within the shock modeled here can be seen in Figure \ref{fig:shock}. It should be noted that there is a distinct delay in the heating of the gas relative to the peak drift velocity and dust temperature (whose calculation will be described in Section \ref{sec:tdust}) by roughly an order of magnitude in time. The impact of this delay will be discussed in Section \ref{sec:p5_results}.

\subsubsection{Dynamic Dust Heating}\label{sec:tdust}
In order to accurately assess the significance of enhancement through desorption versus chemical processing in shocks, it is crucial to consider the effects of the dust temperature on the system. Therefore, we adopted the formalism from \citet{Miura:2017} and \citet{Aota:2015}, which considers the effects from the adiabatic heating of high-velocity particle collisions and the conductive heating from the dust residing in hot gas.  

To start, the prescription for gas-grain energy transfer in the free flow approximation \citep{Probstein:1968,Kitamura:1986} is used to describe the evolution of the dust temperature as a balance between the radiative cooling, collisional heating, and radiative heating.

The temperature of the dust, $T_{\text{dust}}$, can be thus described as
\begin{equation}
	\frac{4}{3} \pi a_{\text{dust}}^2 \rho_{\text{mat}} \frac{\de T_{\text{dust}}}{\de t} = 4\pi a_{\text{dust}}^2 \left( \Gamma - \epsilon_{\text{em}} \sigma_{\text{SB}} T_{\text{dust}}^4 \right) \label{eq:miura9}
\end{equation}
where $a_{\text{dust}}$ is the average dust radius, $\rho_{mat}$ is the mass density within a dust grain, $\epsilon_{\text{em}}$ is the emission coefficient for the dust given as
\begin{equation}
 \epsilon_{em} = A a_{\text{dust}}T_{\text{dust}}^2,
\end{equation}
(where $A$ is the dust emissivity given by \citet{Draine:1984} as $A$=0.122 cm$^{-1}$ K$^{-2}$), $\sigma_{\text{SB}}$ is the Stefan-Boltzmann constant, and $\Gamma$ is the function describing the rate that energy is transferred from the gas to the dust per unit area (i.e. energy flux between gas and dust). Here, $\Gamma$ is given by 
\begin{equation}
	\Gamma = \rho_{\text{gas}} v_{\text{drift}} \left( T_{\text{recovery}} - T_{\text{dust}} \right)C_{\text{H}}
\end{equation}
where $T_{\text{recovery}}$ is the recovery temperature, or adiabatic equilibrium temperature, defined in \citet{Probstein:1968} and \citet{Kitamura:1986} as
\begin{equation}
 T_{\text{recovery}} = \frac{2}{\gamma + 1}\left( \gamma + (\gamma-1)s_a^2 - \frac{\gamma-1}{1 + 2s_a^2 + \frac{s_a}{\sqrt{\pi}e^{s_a^2}\text{erf}(s_a)}} \right)
\end{equation}
and $C_\text{H}$ is the heat transfer function defined by
\begin{equation}
	C_\text{H} = \frac{\gamma+1}{\gamma-1}\frac{k_\text{B}}{8 \mu m_\text{H} s_a^2}\left[ \frac{s_a}{\sqrt{\pi}e^{s_a^2}} + \left(\frac{1}{2} + s_a^2\right)\text{erf}(s_a) \right].
\end{equation}
In these equations, $\gamma$ is the specific heat-ratio, $k_\text{B}$ is the Boltzmann constant, $\mu$ is the mean molecular weight of a gas particle, $m_\text{H}$ is the mass of a hydrogen atom, and $s_a$ is a ratio of the drift velocity and the thermal gas velocity defined by
\begin{equation}
s_a = \frac{v_d}{\sqrt{2k_\text{B} T_{\text{gas}}/\mu m_H}}.
\end{equation}
Equation \ref{eq:miura9} can be solved for the dust temperature at which the heating of the dust by the gas is in equilibrium with the radiative cooling,
\begin{equation}
	\Gamma - 4 \epsilon_{\text{em}} \sigma_{\text{SB}} T_{\text{dust}}^4 = 0,
\end{equation}
which can be further rewritten as
\begin{align}
	C_1 v_{d}^3 \alpha_1(s_a) \alpha_2(s_a) - C_2 T_{\text{dust}} v_d \alpha_2(s_a) - C_3 T_{\text{dust}}^6 &= 0 \\
    T_{\text{dust}}^6 + \frac{C_2}{C_3} v_d \alpha_2 T_{\text{dust}}- \frac{C_1}{C_3} v_d^3 \alpha_1 \alpha_2 &= 0 \label{eq:td_reduced}
\end{align}
where the $C$ values are constants with respect to $s_a$ defined by
\begin{align}
C_1 &= \frac{1}{8} \rho_{\text{gas,0}}\\
C_2 &= \frac{1}{8}\frac{\gamma+1}{\gamma-1} \frac{k_\text{B}}{\mu m_\text{H}} \rho_{\text{gas,0}} \\
C_3 &= A \sigma_{\text{SB}} a_{\text{dust}}
\end{align}
and $\alpha$ values are functions of $s_a$ defined by
\begin{align}
	\alpha_1(s_a) &= 1 + \frac{\gamma}{\gamma-1} s_a^{-2} - \frac{1}{1+2s_a^{-2} + 2 \frac{s_a}{\sqrt{\pi}e^{s_a^2}\text{erf}(s_a)} } \\
    \alpha_2(s_a) &= \left(1 + \frac{1}{2s_a^2}\right)\text{erf}(s_a) + \frac{1}{\sqrt{\pi}s_a e^{s_a^2}}
\end{align}
With this reduction of the equation, Equation \ref{eq:td_reduced} is solved for $T_{\text{dust}}$ using the \textsc{minpack}'s \textsc{hybrd} and \textsc{hybrj} routines with the modified Powell method \citep{More:1980dr}. For the purposes of guiding the algorithm, initial guesses were computed by
\begin{equation}
	T_{\text{dust,guess}} = T_{\text{dust,0}}  + \frac{v_d}{v_s}\left(10 \text{ K} \right).
\end{equation}
Applying the temporal evolution of the gas temperature, density, and drift velocity determined in the previous section, we are able to compute the dust temperature across the evolution of the shock. In the pre-shock gas, the gas and dust are assumed to be in thermal equilibrium, or $T_{\text{dust}} = T_{\text{gas}}$.

In addition, we deviate from the formalism used in \citet{jimenezserra:2008}, by setting an additional constraint of the velocity when used to calculate the rate of dust heating. Specifically, we assume that as the gas and grains begin to re-couple, the drift velocity will eventually fall below the random thermal velocities of the gas. In the hot-gas regime, this thermal velocity may be non trivial. As such, the velocity of collisions may be more accurately described for these purposes as:
\begin{equation}
	v = \text{max}\left[ v_d,\sqrt{\frac{2 k_B T_{\text{gas}}}{\mu m_\text{H}}} \right].
\end{equation}

As can be seen in Figure \ref{fig:shock}, the grain temperature is found to initially track with the velocity structure within the shock, peaking at the same corresponding time as the peak velocity due to adiabatic heating. However, as the gas continues to heat up, the grain temperature undergoes a secondary heating regime due to the conductive heating from the hot gas. As such, while the sputtering rate (see Section \ref{sec:sputtering}) will fall off rather rapidly, the ability for species to redeposit onto the grains is delayed until the gas can cool sufficiently.

\subsubsection{Sputtering} \label{sec:sputtering}
One major physical process that is common in astrophysical shocks is the non-thermal desorption of molecules from the ice into the gas phase through the collision of moderately fast gas-phase particles with the dust grains (e.g. sputtering). This process is a key mechanism for lifting complex molecules into the gas phase non-destructively. For typical C-shock velocities (e.g. 10-40 km s$^{-1}$), the sputtering is significant enough to desorb most  species from the ice mantle without destroying them or having significant grain core erosion, both of which are common in high-velocity J-shocks \citep{RequenaTorres:2006ki}.

In order to include these processes, the sputtering yields must be computed for each species on the ice, which is dependent on the conditions of the shock, the nature of the incident particle, and the binding properties of the target molecule. Utilizing the procedure discussed in \citet{Caselli:1997} and \citet{jimenezserra:2008}, with the initial formalism introduced in \citet{DraineSalpeter:1979}, we can compute the rate at which some ice species, $J$, is sputtered off the surface by some incident particle, $P$, such that
\begin{align}
	\left[\frac{\de n_{\text{J}}}{\de t}\right] = &\frac{\pi a_{\text{dust}}^2 n_P}{2s_P} \sqrt{\frac{8k_B T_{\text{gas}}}{\pi m_P}} \\
	&\int_{x_{\text{th}}}^{\infty} x^2 \left(e^{-(x-s_P)^2} - e^{-(x+s_P)^2} \right) \left< Y(E_P) \right> \de x
\end{align}
where $a_{\text{dust}}$ is the average dust grain radius, $n_P$ is the gas-phase density of the projectile, $k_B$ is the Boltzmann-constant, $T_{\text{gas}}$ is the gas temperature, $m_P$ is the mass of the projectile. $s_P$ is the ratio between the kinetic energy of a particle traveling at the current drift velocity and the thermal energy of the gas, or
\begin{equation}
	s_P^2 \equiv \frac{m_P v_d^2}{2k_B T_{\text{gas}}}. 
\end{equation}
Similarly, $x$ is the ratio between the kinetic energy of a projectile at some specific velocity, $v_x$, and the thermal energy of the gas. This is related to the impact energy of the projectile, $E_P$ as 
\begin{equation}
	E_P = x^2 k_B T_{\text{gas}}.
\end{equation}
The integral, therefore, is equivalent to the total contribution from projectiles at all velocities corresponding to an $x$ value higher than $x_{\text{th}}$, which is defined as 
\begin{equation}
   x_{\text{th}} = \sqrt{\frac{\epsilon_0 U_0}{\eta k_B T_{\text{gas}}}}
\end{equation}
where $U_0$ is the binding energy of the species to be sputtered off and $\epsilon_0$ is defined as
\begin{equation}
	\epsilon_0 = \text{max}[1,4\eta].
\end{equation}
Here, $\eta$ is a ratio describing the relative masses of the projectile ($m_P$) and the target species ($m_T$)
\begin{equation}
	\eta = 4 \xi \frac{m_P m_T}{(m_P + m_T)^{-2}}
\end{equation}
where $\xi$ is a material-specific efficiency factored assumed to be 0.8 for ices \citep{DraineSalpeter:1979}.
The angle-averaged value for the sputtering yield, $Y$, which is approximately equal to double the perpendicular yield value, is described in \citet{DraineSalpeter:1979} such that
\begin{equation}
	\left< Y \right>_{\theta} \approx 2Y(\theta =0) = K \frac{(\epsilon - \epsilon_0)^2}{1 + \left(\frac{\epsilon}{30}\right)^{4/3}}  \text{ for } \epsilon > \epsilon_0
\end{equation}
where $K\approx$8.3$\times$10$^{-4}$ and $\epsilon$ is defined as
\begin{equation}
	\epsilon = \frac{E_P}{U_0} \eta.
\end{equation}
For the purposes of calculating these sputtering yields, the binding energies, $U_0$, need to be known for each species; these are already given within \textsc{nautilus} as part of the \textsc{KIDA 2014} network. It should be noted that values for many of these species are estimated and have not been explicitly experimentally measured.

For each grain species in the network, a sputtering reaction was added for the following gas-phase projectiles: H$_2$, He, C, O, Si, Fe, and CO. To compute $n_P$, the gas-phase abundance within the model at each time step for each projectile was multiplied by the overall gas density (i.e. $n_{\text{H}_2}$). In the \textsc{nautilus} three-phase model, sputtering can only occur for species on the surface of the ice (i.e. the top few monolayers), similar to the vast majority of other desorption mechanisms \citep{Ruaud:2016bv}. The integral over $x$ was computed with the \textsc{dqags} routine in the \textsc{quadpack} package\footnote{http://www.netlib.org/quadpack/}
. Computing this indefinite integral numerically was found to be highly sensitive to the sampling rate of the integrating routine, as much of the range in which the integral is evaluated over is trivially small. As such, to increase the efficiency of the computation of the integral, we converted the integral to be definite by approximating the upper and lower bounds such that:
\begin{itemize}
	\item If $x_{\text{th}} > s_P + 20$, then the vast majority of the projectiles at this drift velocity are traveling at too low of a velocity to effectively sputter. In this case, we assume the integral will evaluate to zero.
    \item If $x_{\text{th}} < s_P - 10$, then the vast majority of the projectiles at this drift velocity are potentially good projectiles for efficient sputtering. In this case, we reset the lower bound of the integral to $s_P - 10$ and the upper bound to $s_P + 20$.
    \item Otherwise, $x_{\text{th}}$ is reasonably close to $s_P$ and will not cause any significant issues integrating as a lower bound. In this case, we only reset the upper bound to $s_P + 20$ for computational efficiency instead of evaluating an indefinite integral. 
\end{itemize}

\subsection{High Temperature Chemical Network} \label{sec:hitemprx}
One major motivation for the use of the latest version of the \textsc{osu.kida} network is the inclusion of the high-temperature reactions developed by \citet{Harada:2010}, as the gas-phase temperatures in the post-shock regimes can reach 1000 K at even $v_s\sim$20 km s$^{-1}$. Since many gas-grain chemical networks are optimized to reproduce the chemistry of dark clouds \citep{Hincelin:2011fr,shingledecker_general_2018,Loomis:2016jsa,mcguire_detection_2017,mcguire_detection_2018} and hot cores\citep{Garrod:2013id,hincelin_new_2015}, this high temperature regime has remained insufficiently studied. By considering both the dark cloud and high-temperature regimes in the same model, we can evaluate how effective the network is under a large dynamic range of conditions. 

In addition, we also added several gas and grain reactions for species related to the common observational shock-tracer, HNCO. The network for HNCO-isomers and related ions developed by \citet{Quan:2010ba} was included in order to fully account for the potential of multiple formation routes, especially in colder conditions. Moreover, in order to test the importance of the unique chemistry occurring in the high-temperature regime, we added several reactions, listed in Table \ref{tab:tsang_reactions}, with significant barriers and a strong temperature dependence taken from previous combustion chemistry studies \citep{tsang_chemical_1991,tsang:1992}.

\begin{table}
\centering
\scriptsize
\caption{Reactions added to network from \citet{tsang_chemical_1991,tsang:1992}. }
\begin{tabular}{@{\hskip3pt}c@{\hskip3pt} c c c @{\hskip1pt}c@{\hskip1pt}}
\hline\hline
Reactions				&		$\alpha$			&	$\beta$	&	$\gamma$  &  Temperature Range	\\
					&[cm$^{3}$ s$^{-1}$]	&				&	[K] & [K]	\\
\hline
\ce{H2}+OCN$\rightarrow$HNCO+H   &6.54$\times$10$^{-14}$ &  2.58 &   2720 & 300-3300\\
\ce{H2O}+OCN$\rightarrow$HNCO+OH & 9.13$\times$10$^{-14}$ & 2.17 &  3050 &  500-2500\\
HCN+OH$\rightarrow$HNCO+H & 4.18$\times$10$^{-18}$ & 4.17 & -247.8 & 298-2840\\
HCN+OH$\rightarrow$HOCN+H & 6.11$\times$10$^{-14}$ & 2.45 & 6100 &  298-2840\\
OH+OCN$\rightarrow$HNCO+O & 5.38$\times$10$^{-14}$ & 2.27 & 496.7 & 500-2500\\
\end{tabular}
\label{tab:tsang_reactions}
\end{table}




\section{Results} \label{sec:p5_results}

\subsection{Overview of shock-chemistry regimes}
From the physical structure described in Section \ref{sec:shockstructure}, six major regimes of chemistry in the shock can be seen in the molecular abundances within the model. First, the pre-shock chemistry sets the initial gas and ice abundances after the first 10$^6$ years, representing how the chemistry would proceed in the region in the absence of a shock. Second, the drift velocity begins to rise. Before the sputtering becomes highly efficient in sublimating the ice mantle, the lower-velocity projectile particles are only capable of heating the dust. As such, the mobility of species in the ice rapidly increases, promoting the rapid formation of complex chemistry on the ice for $\sim$50 years, after which the peak amount of sputtering, which occurs up until about 100 years after the shock, lifts essentially all the ice-residing molecules into the gas phase. Following this, the gas, which is heated more slowly than the dust, reaches a peak temperature of $\sim$1000 K around 10$^3$ years after the onset of the shock. At the same time, the dust heating becomes dominated by conductive heating over adiabatic heating as the drift velocity decreases - at which point, the hot gas-phase chemistry becomes much more significant. We refer to this regime as ``post-shock chemistry.'' Previously studies have shown that species can be grouped by how much they are enhanced or destroyed by this post-shock chemistry \citep{Collings:2004jp,Holdship:2017}.

After the environment cools back down to dark cloud temperatures, fully returning around 10$^4$ years post-shock, the species in the gas phase redeposit back onto the grain and reform the ice. Hypothetically, then the structure and composition of ice in the post-shock regime might be distinct from ice formed under quiescent cold-core conditions. Thus, it is possible that there will be permanent relics of a shock occurring in ice. Reactions thus stimulated on the surface of this newly enriched ice also may result in an increased rate of chemical desorption, i.e. where a reaction on the ice surface is energetic enough to lift the products into the gas-phase \citep{garrod_non-thermal_2007}. This will be discussed in detail in Section \ref{sec:othershockprobes}. Following this period, we continue to run the model until 10$^6$ years where any physical, non-chemical relics of the shock should be no longer relevant.

\subsection{Bulk Evolution of the Ice} 
To follow the overall development of the ice, it is useful to look at one of its primary constituents: H$_2$O. As seen in the left plot in Figure \ref{fig:modeloutline}, the first 10$^6$ years actively builds up a thick ice (typically $\sim$70 monolayers) in the dark cloud conditions prior to the shock. The second 10$^6$ years (i.e. right plot) in Figure \ref{fig:modeloutline} is entirely where the shock occurs. Due to difficulties in viewing data in log-space, we will henceforth exclusively display the abundance evolution from the post-shock time and not the initial time of the model. For the purposes of viewing the results of the pre-shock chemistry, we will only consider the abundance at $t_{\text{postshock}}=0$ or $t_{\text{total}}$=10$^6$ years. In general, following the initial cold core phase, most species tend to reach abundances that do not vary over small changes to the pre-shock timescale. 

Initially, as the drift velocity increases, increased mobility on the heated dust can induce temporary enhancements in the ice abundance for certain species. In the peak sputtering regime, the ice-surface and ice-mantle abundances are seen to drop to trivial levels, with a slight enhancement of the gas-phase abundance of CO. Once the drift velocity decreases to levels that make sputtering inefficient, the ice abundances stop decreasing. However, because the dust is conductively heated by the hot gas in this later regime, the recently sputtered species are unable to redeposit back onto the surface of the ice and will undergo additional post-shock gas-phase chemistry if efficient routes exist. While this regime does not appear to impact the abundance of gas-phase \ce{H2O} in Figure \ref{fig:modeloutline}, specifically, the regime is still relevant for the evolution of the ice and to less abundant, and generally more complex, molecules.

Following the redeposition, the higher post-shock gas density slightly enhances the abundance of some species, but overall should return to initial physical conditions. As will be discussed, some species undergo an enhancement of ice abundances long after the shock passes, despite a gas-phase abundance that is consistent with the pre-shock values. For other species, the ice abundance, with no significant formation routes in dark clouds, will return to the pre-shock value, though only on the timescales of 10$^6$ years after the shock event.



\subsection{Classification of Species}
\begin{figure*}[tb]
\centering     
\subfigure[\ce{CO}]{\label{fig:co}\includegraphics[width=0.45\textwidth]{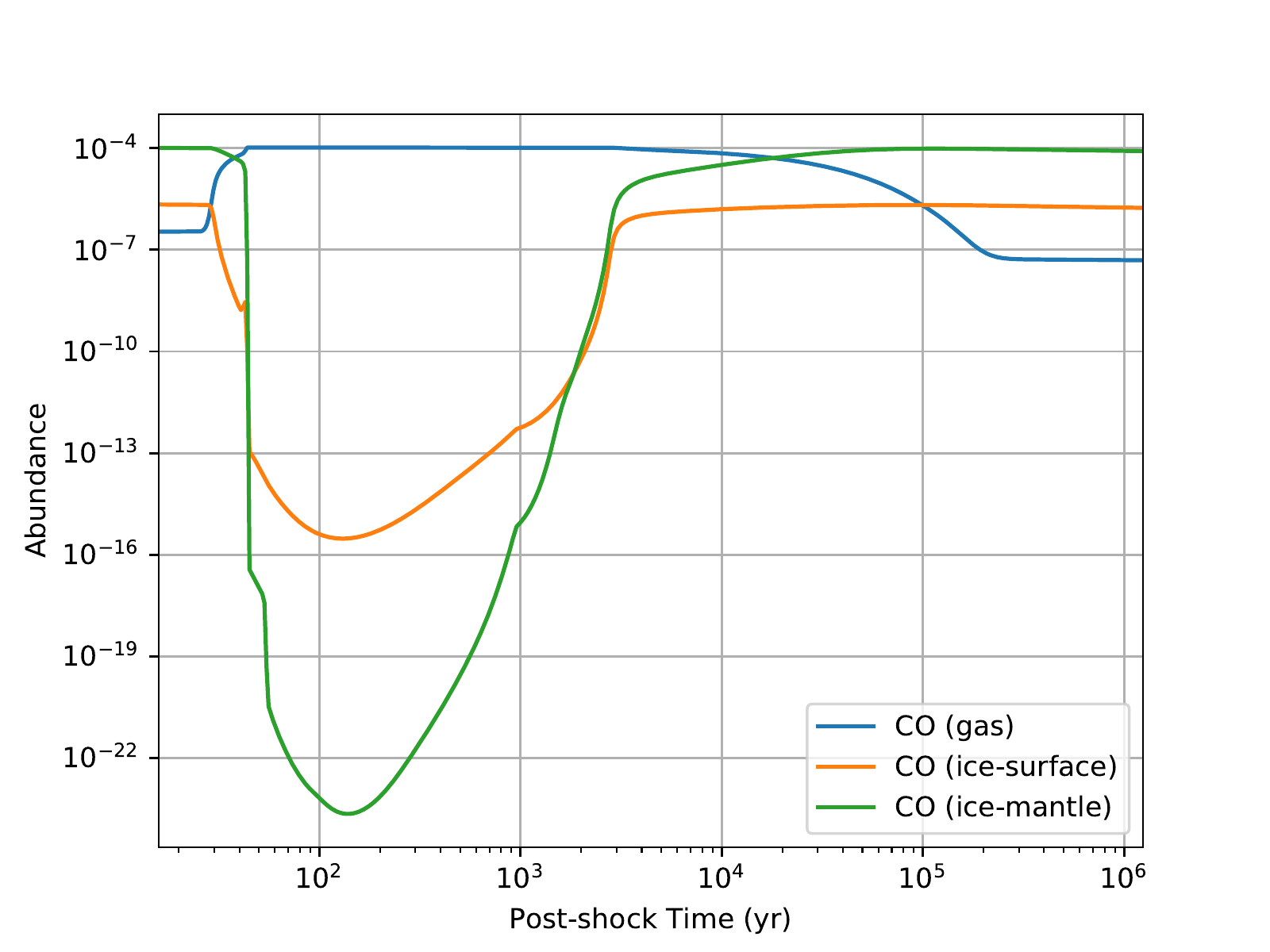}}
\subfigure[\ce{H2O}]{\label{fig:h2o }\includegraphics[width=0.45\textwidth]{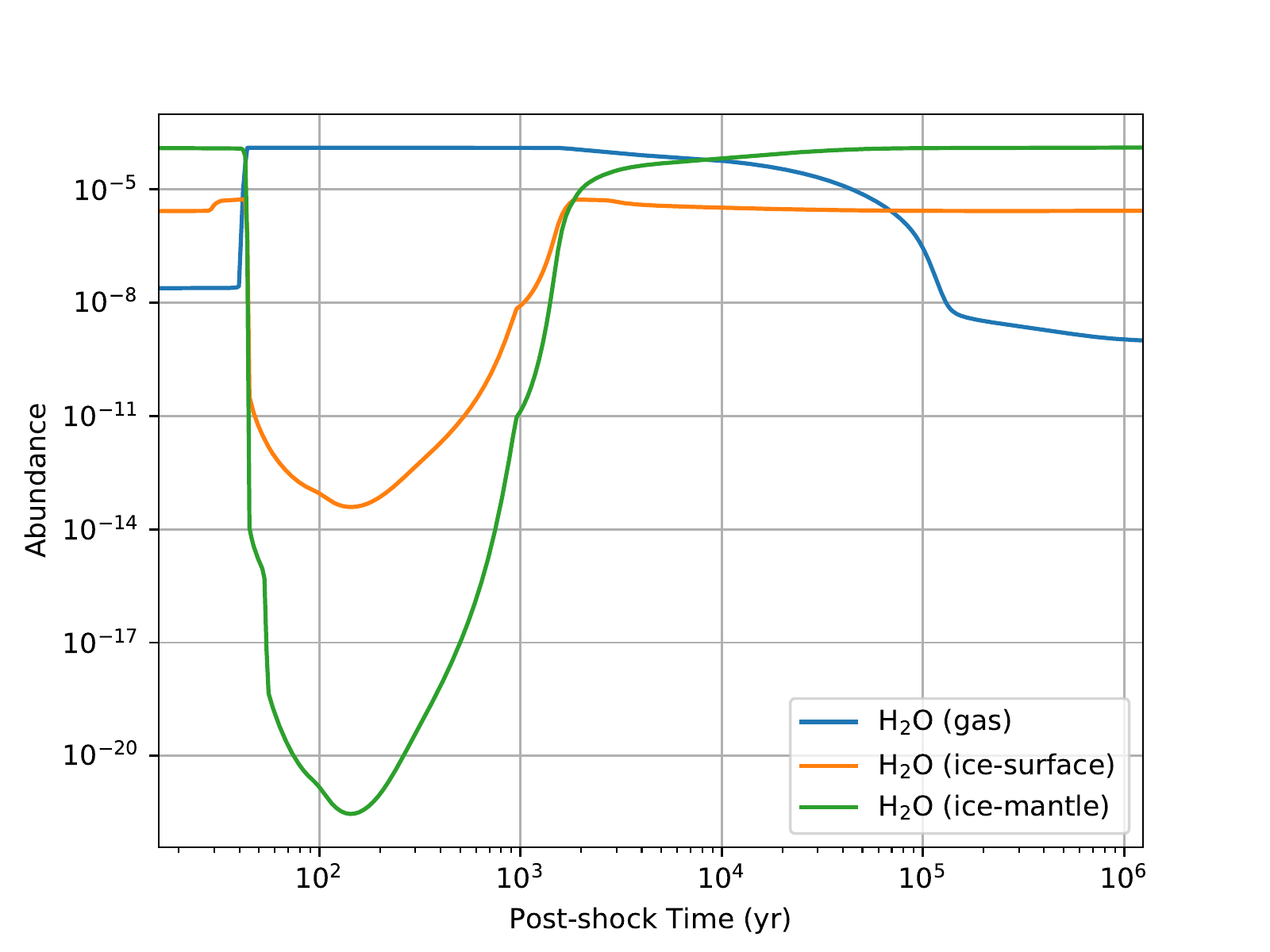}}
\subfigure[\ce{NH3}]{\label{fig:nh3}\includegraphics[width=0.45\textwidth]{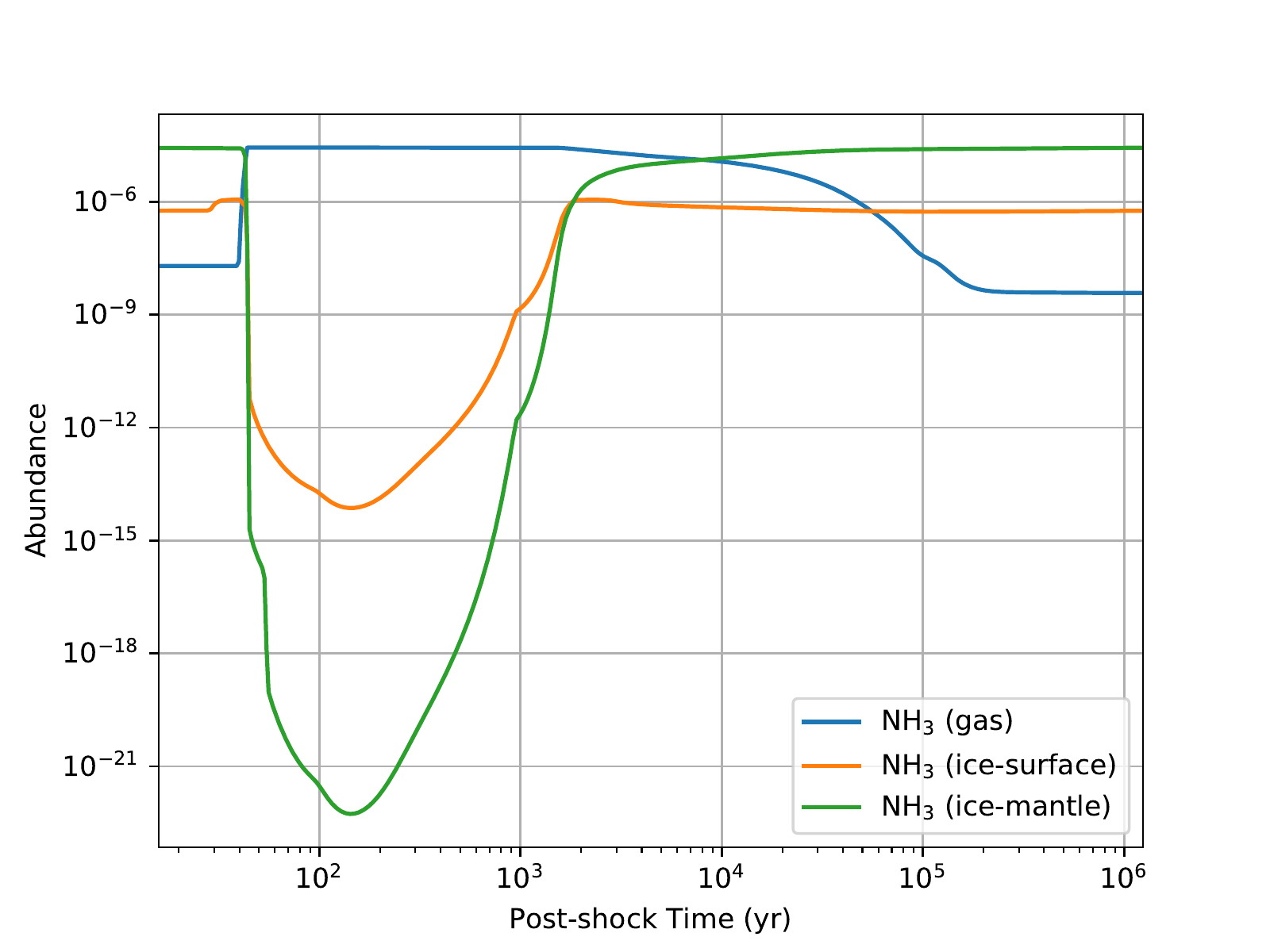}}
\subfigure[\ce{HCN}]{\label{fig:hcn}\includegraphics[width=0.45\textwidth]{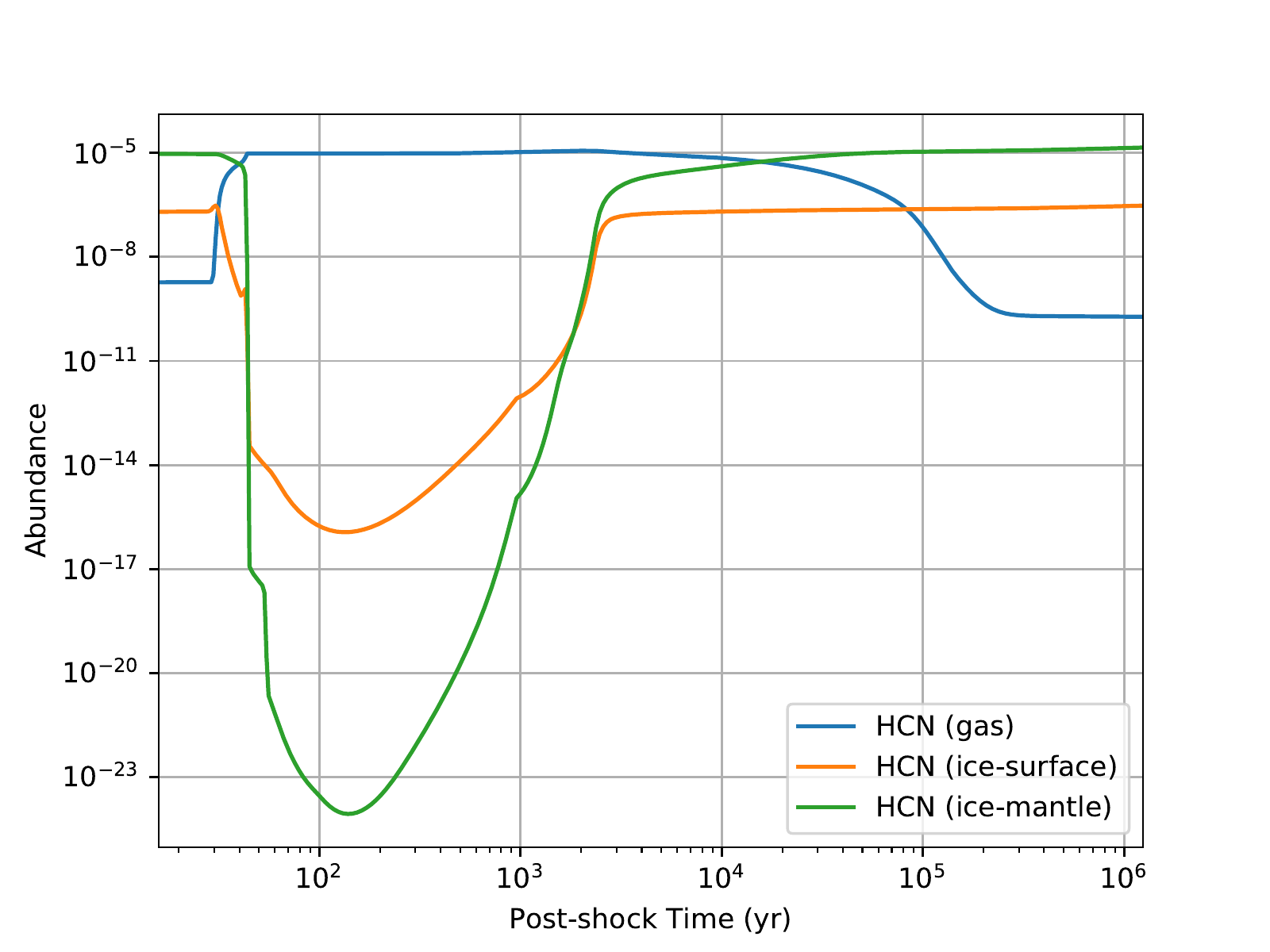}}
\caption{Simulated three-phase abundances of studied species over the time period in which the shock passes through in the model (i.e. $t_{\text{postshock}}$=0 corresponds to $t_{\text{total}}$=10$^6$ years) for species with little post-shock chemistry and whose abundance traces the bulk evolution of the ice.}
\label{fig:3phase1}
\end{figure*}

\begin{figure*}[tb]
\centering     
\subfigure[\ce{CH3OH}]{\label{fig:ch3oh}\includegraphics[width=0.45\textwidth]{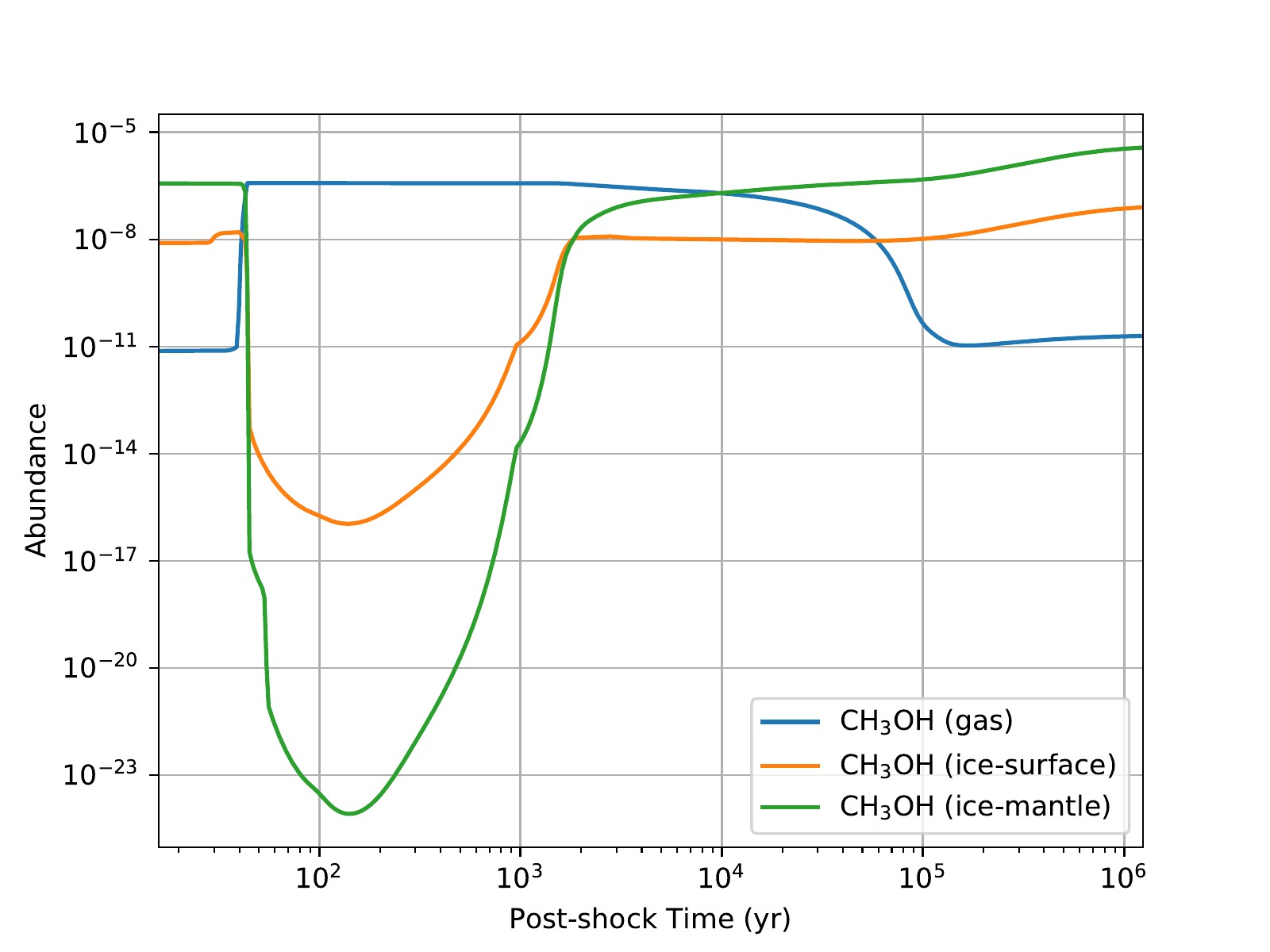}}
\subfigure[\ce{H2CO}]{\label{fig:h2co}\includegraphics[width=0.45\textwidth]{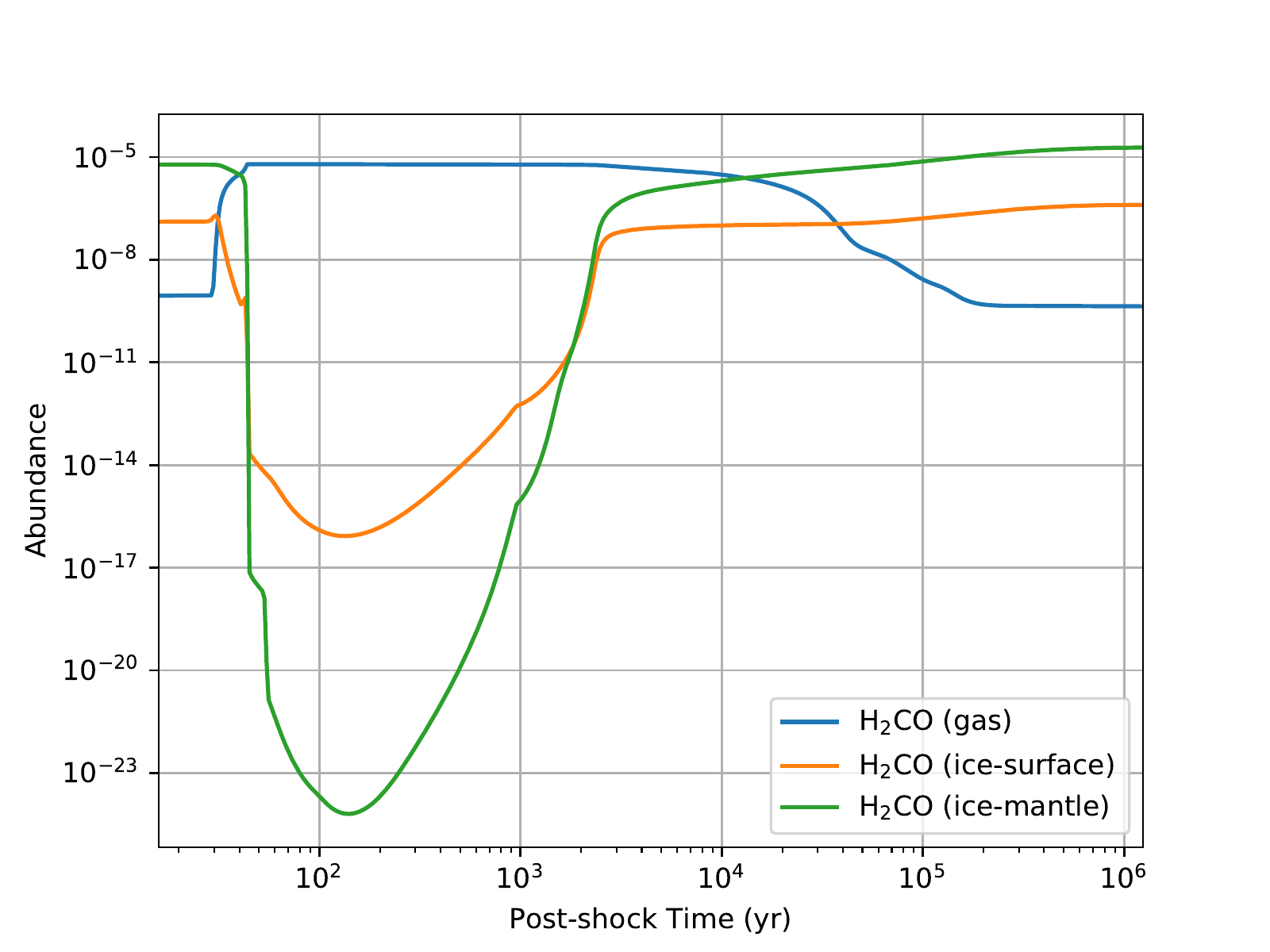}}
%
\subfigure[\ce{HCOOH}]{\label{fig:hcooh}\includegraphics[width=0.45\textwidth]{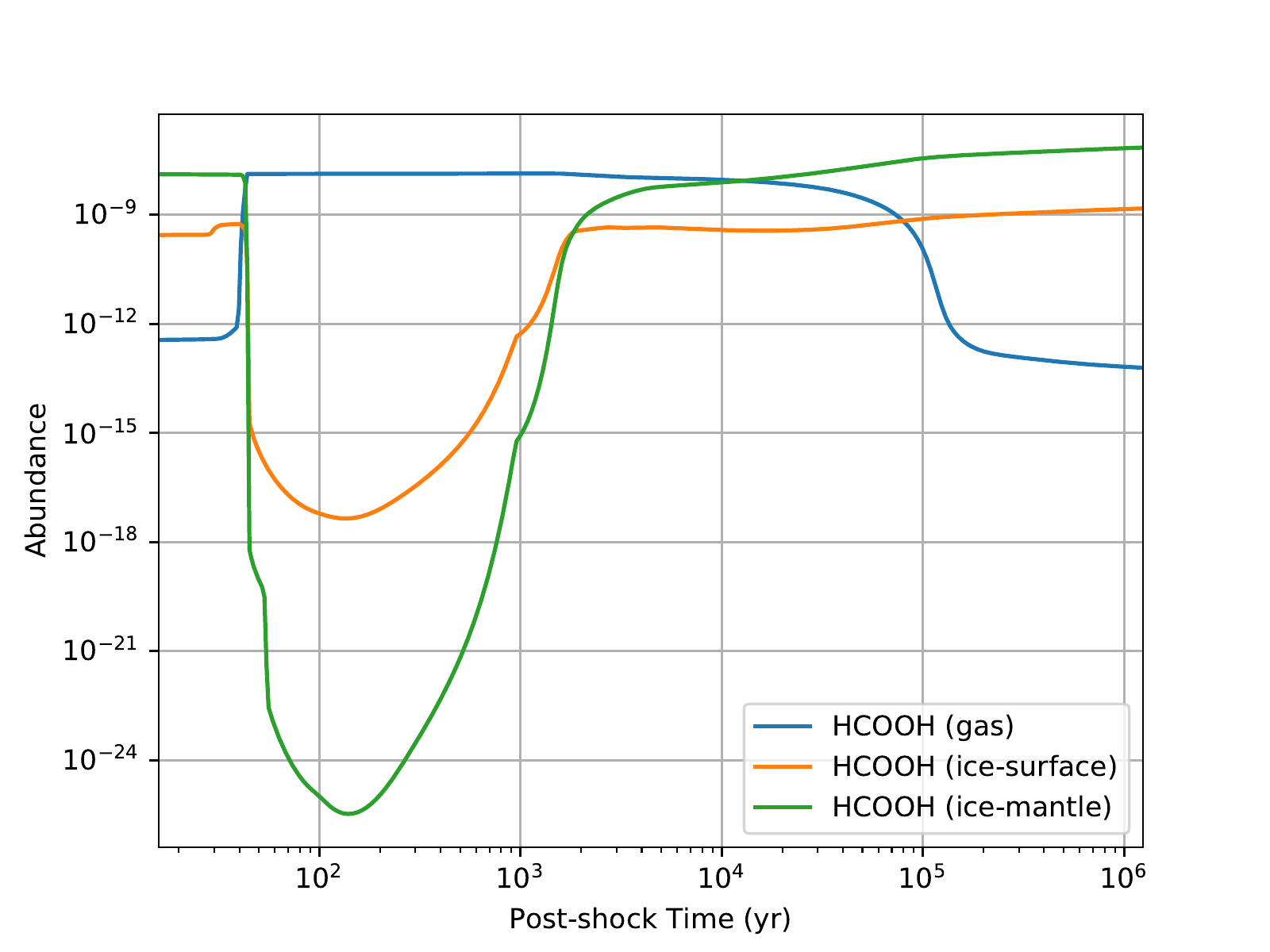}}
\subfigure[\ce{CH3OCH3}]{\label{ch3och3}\includegraphics[width=0.45\textwidth]{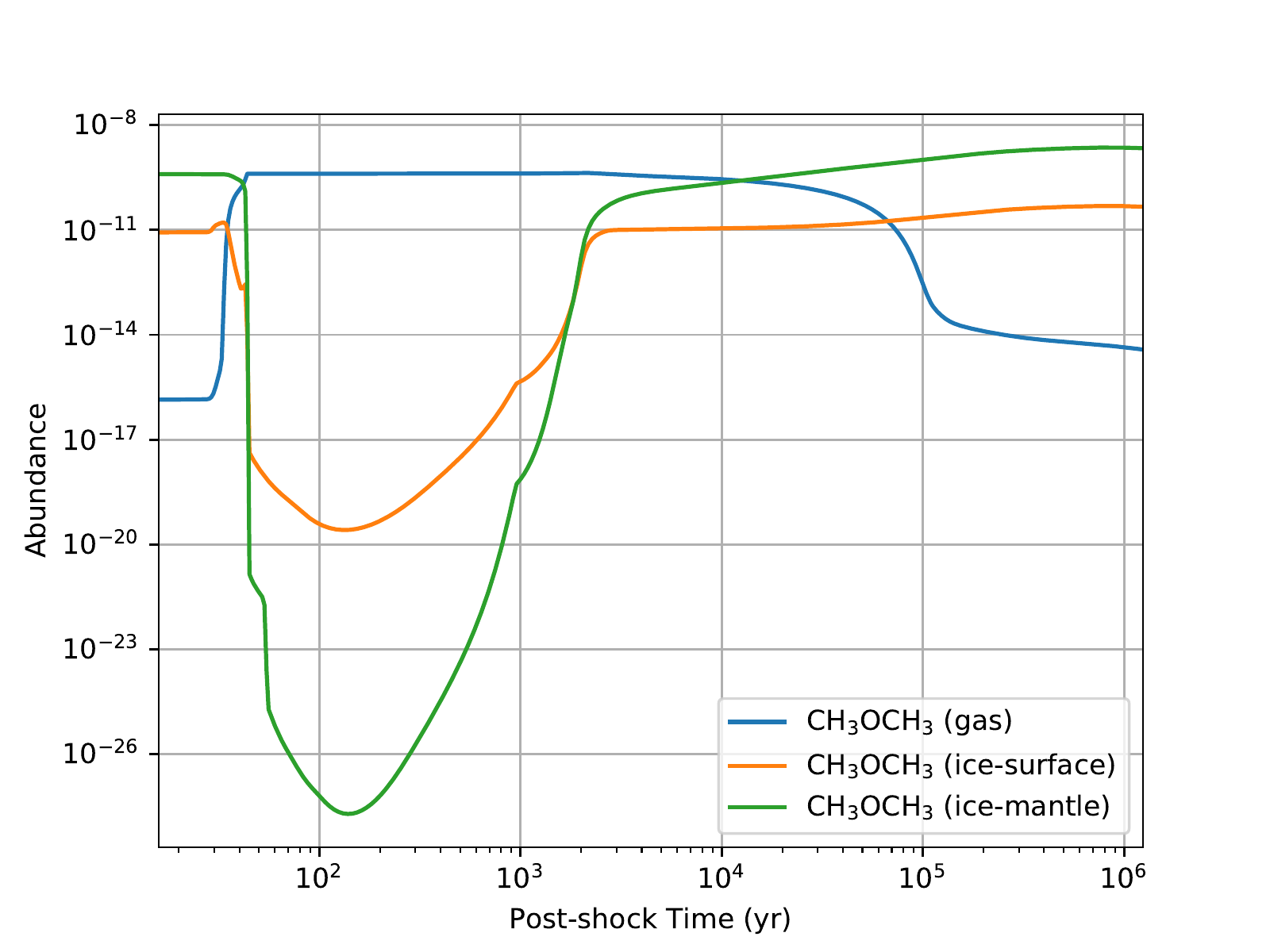}}
\caption{Simulated three-phase abundances of studied species over the time period in which the shock passes through in the model (i.e. $t_{\text{postshock}}$=0 corresponds to $t_{\text{total}}$=10$^6$ years) for species with little post-shock chemistry and whose abundance traces the bulk evolution of the ice.}
\label{fig:3phase4}
\end{figure*}

\begin{figure*}[tb]
\centering     
\subfigure[\ce{NH2CHO}]{\label{fig:nh2cho}\includegraphics[width=0.47\textwidth]{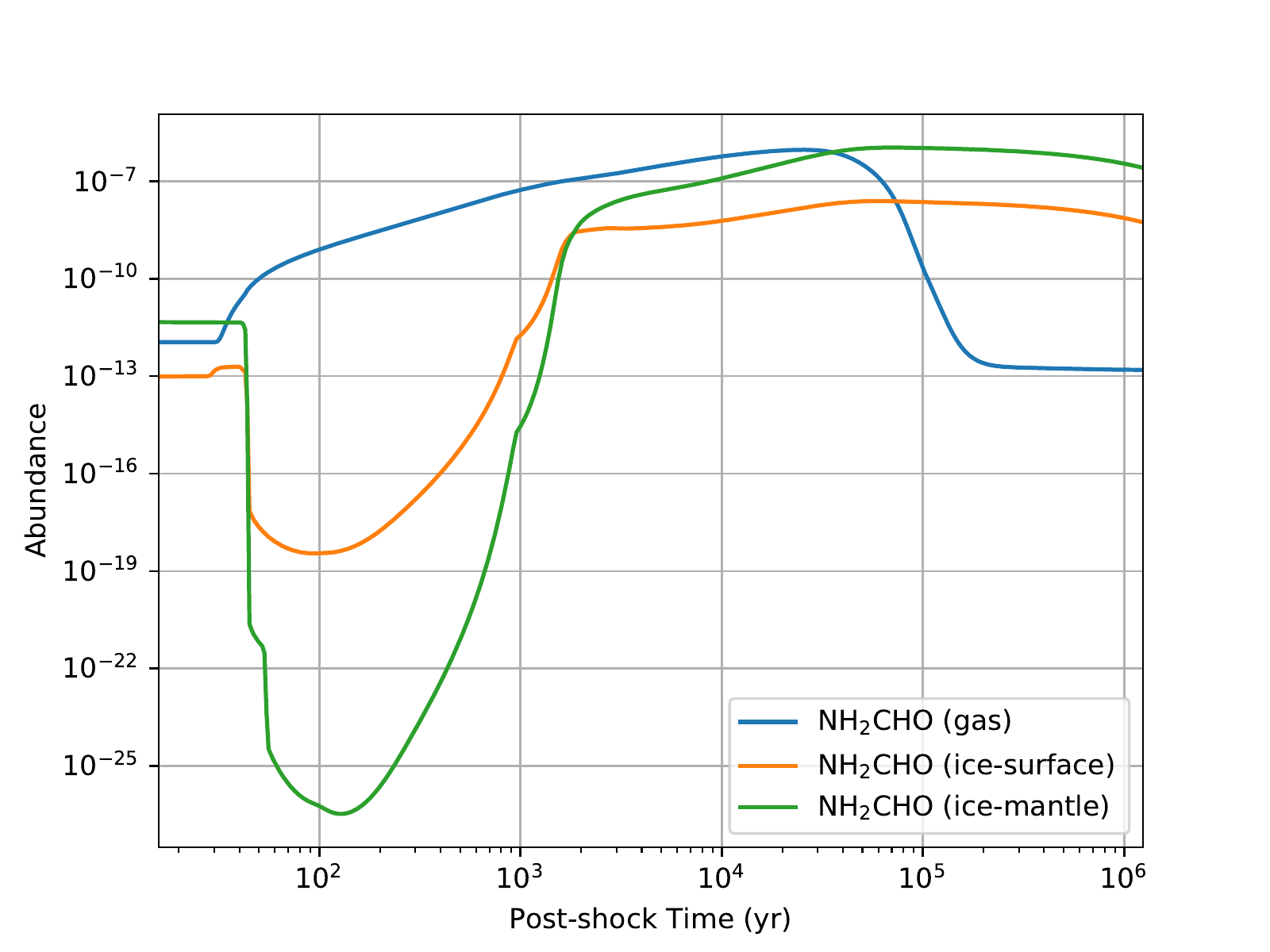}}
\subfigure[\ce{HCOOCH3}]{\label{fig:hcooch3}\includegraphics[width=0.47\textwidth]{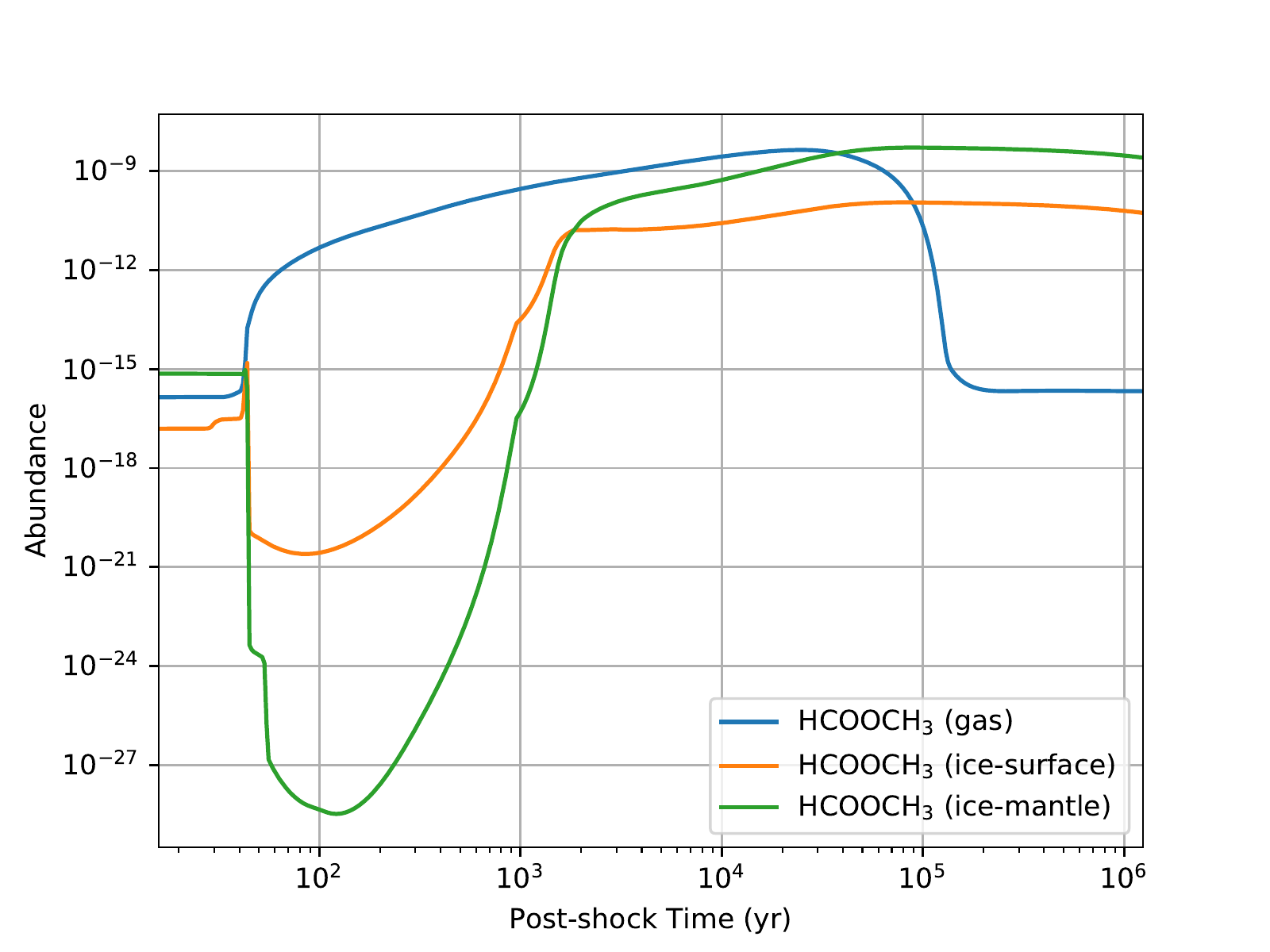}}
%
\subfigure[\ce{CH3CHO}]{\label{fig:ch3cho}\includegraphics[width=0.45\textwidth]{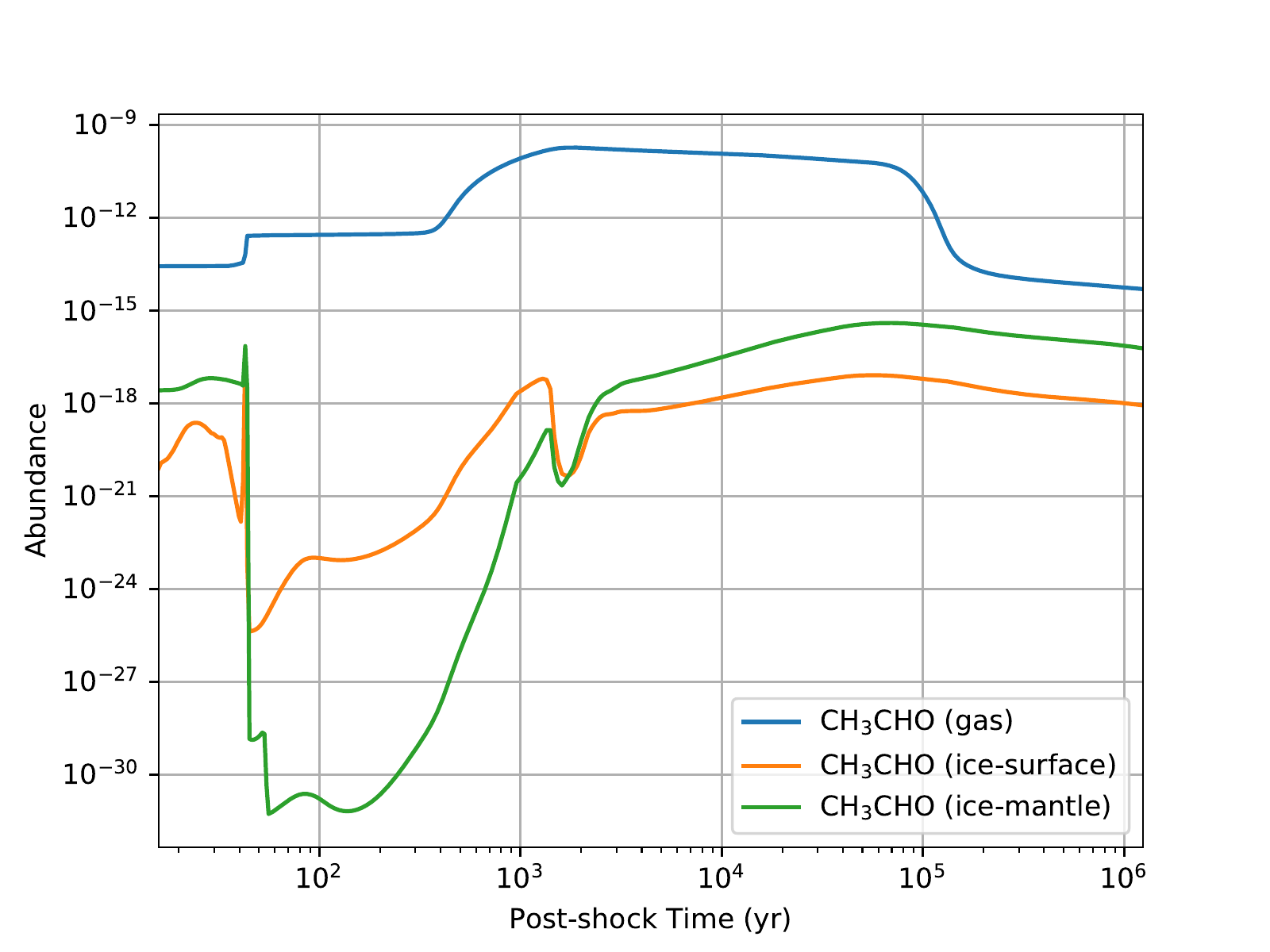}}
\caption{Simulated three-phase abundances of studied species over the time period in which the shock passes through in the model (i.e. $t_{\text{postshock}}$=0 corresponds to $t_{\text{total}}$=10$^6$ years) for species that display discrete gas-phase enhancement through post-shock chemistry.}
\label{fig:3phase6}
\end{figure*}

\begin{figure*}
\centering     
\subfigure[\ce{CS}]{\label{fig:cs}\includegraphics[width=0.45\textwidth]{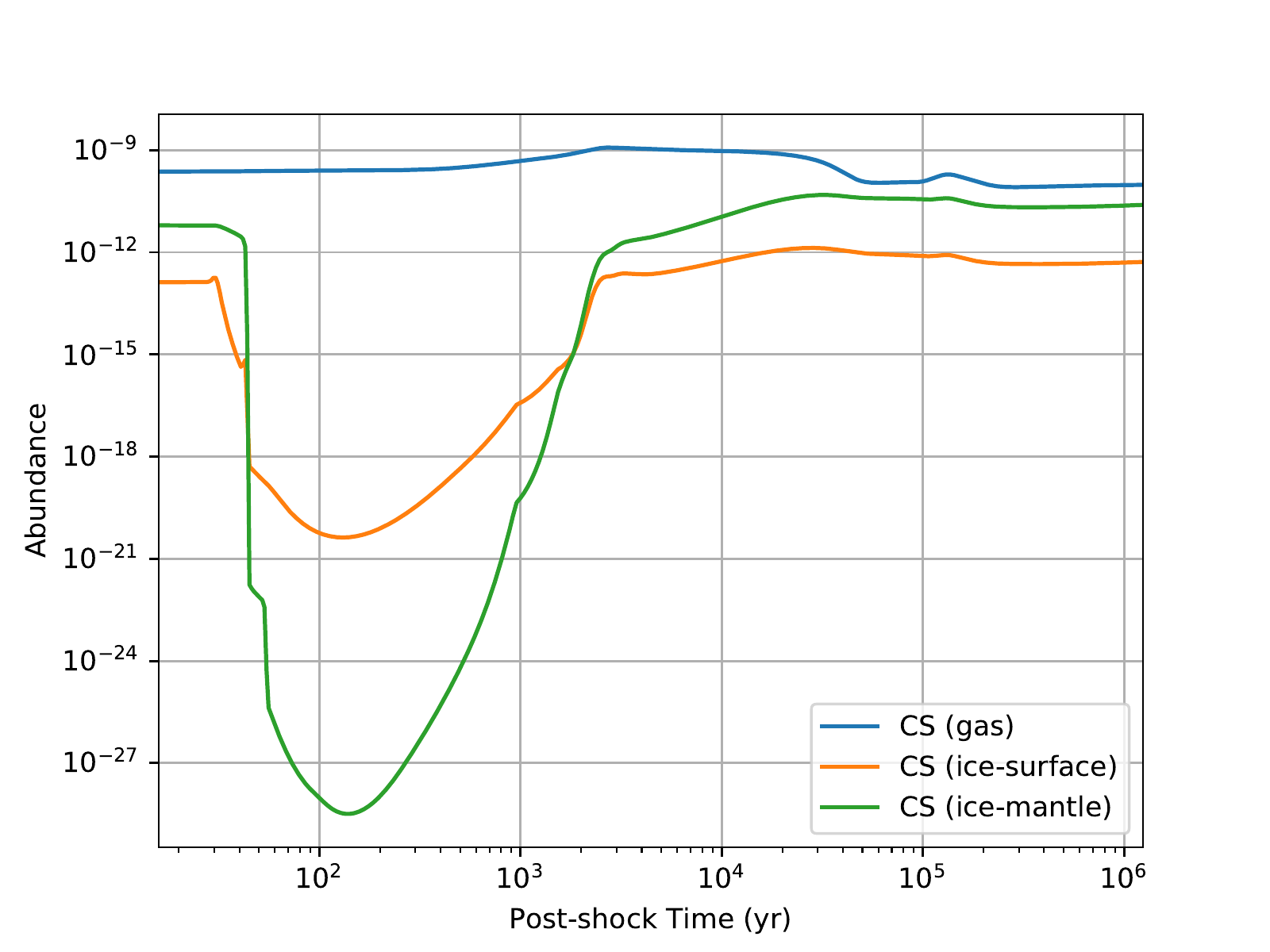}}
\subfigure[\ce{NO}]{\label{fig:no}\includegraphics[width=0.45\textwidth]{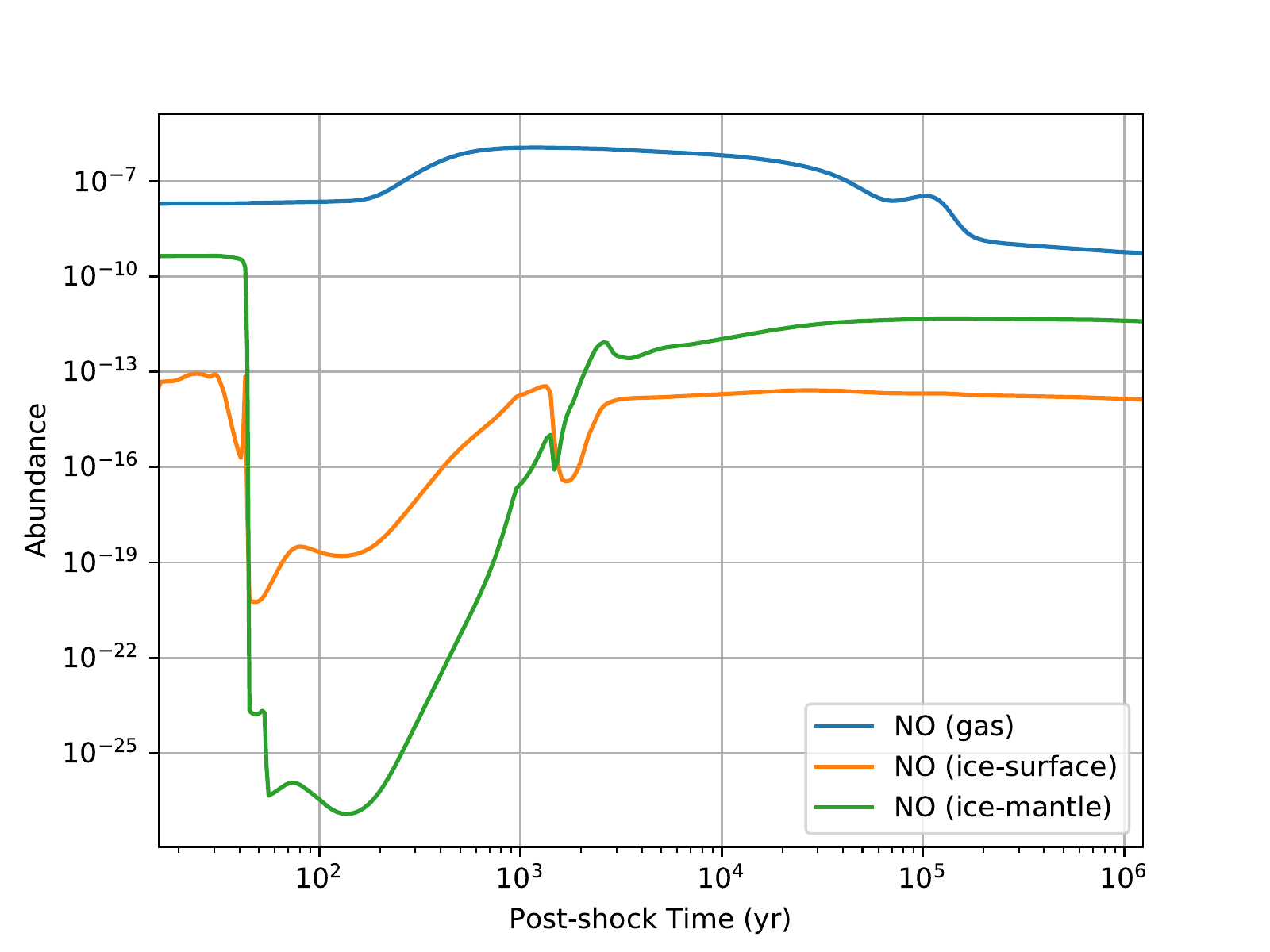}}
\caption{Simulated three-phase abundances of studied species over the time period in which the shock passes through in the model (i.e. $t_{\text{postshock}}$=0 corresponds to $t_{\text{total}}$=10$^6$ years) for species that display reasonably-strong gas-phase enhancements during the post-shock phase.}
\label{fig:3phase7}
\end{figure*}

\begin{figure*}[tb]
\centering     
\subfigure[\ce{HNCO}]{\label{fig:hnco}\includegraphics[width=0.45\textwidth]{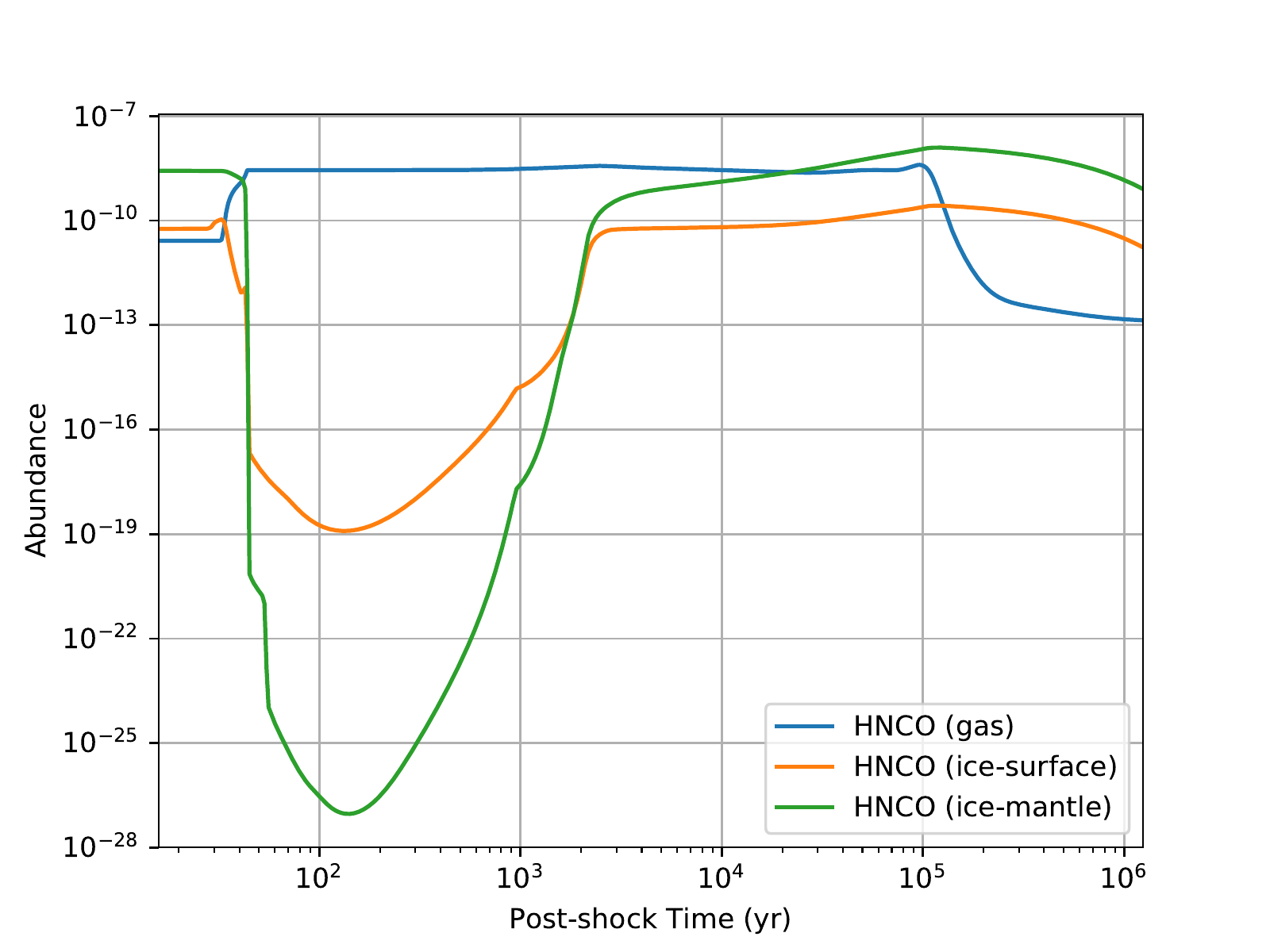}}
\subfigure[\ce{CH3COCH3}]{\label{fig:chs3och3}\includegraphics[width=0.45\textwidth]{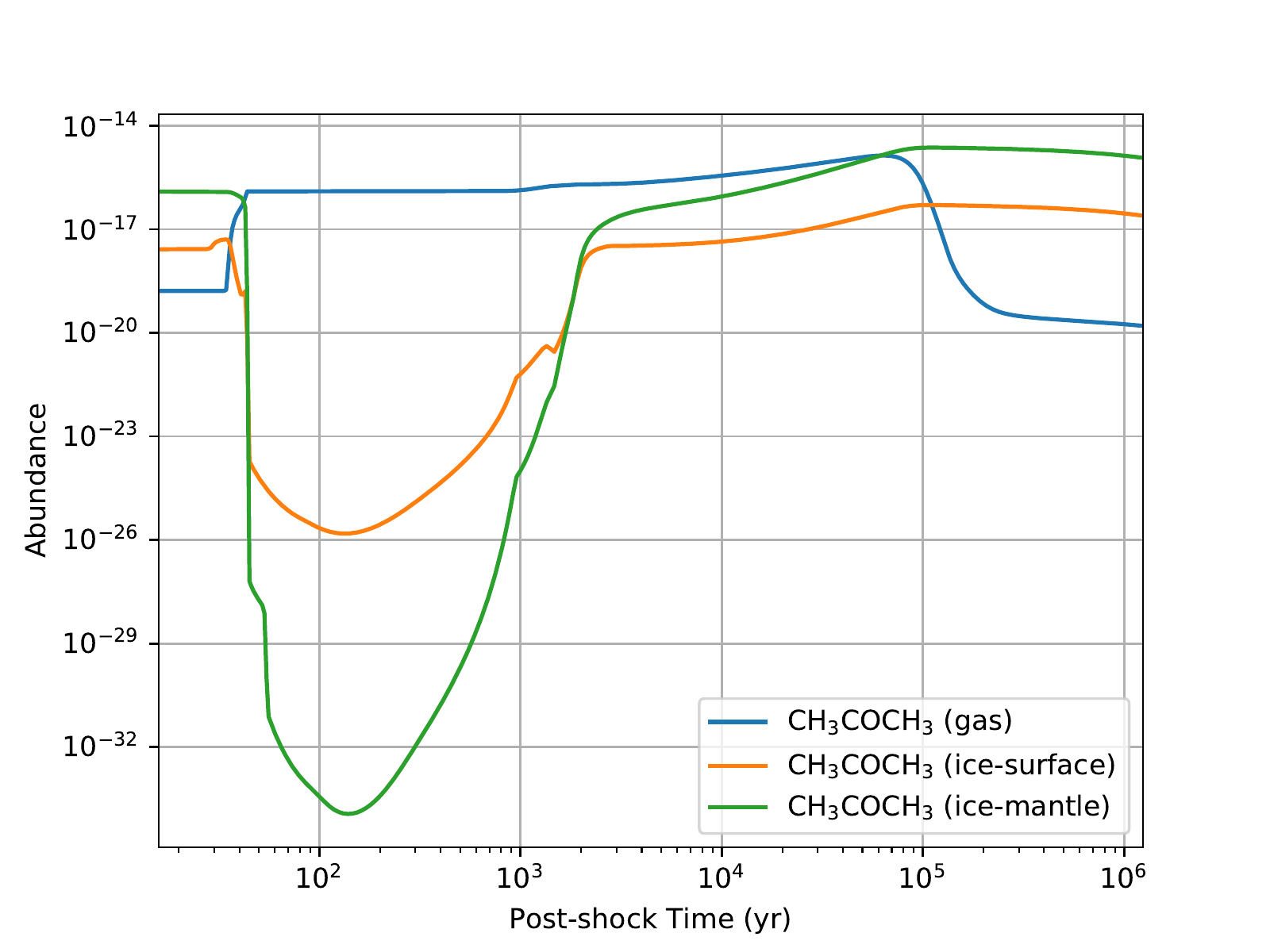}}
\subfigure[\ce{SO}]{\label{fig:so}\includegraphics[width=0.45\textwidth]{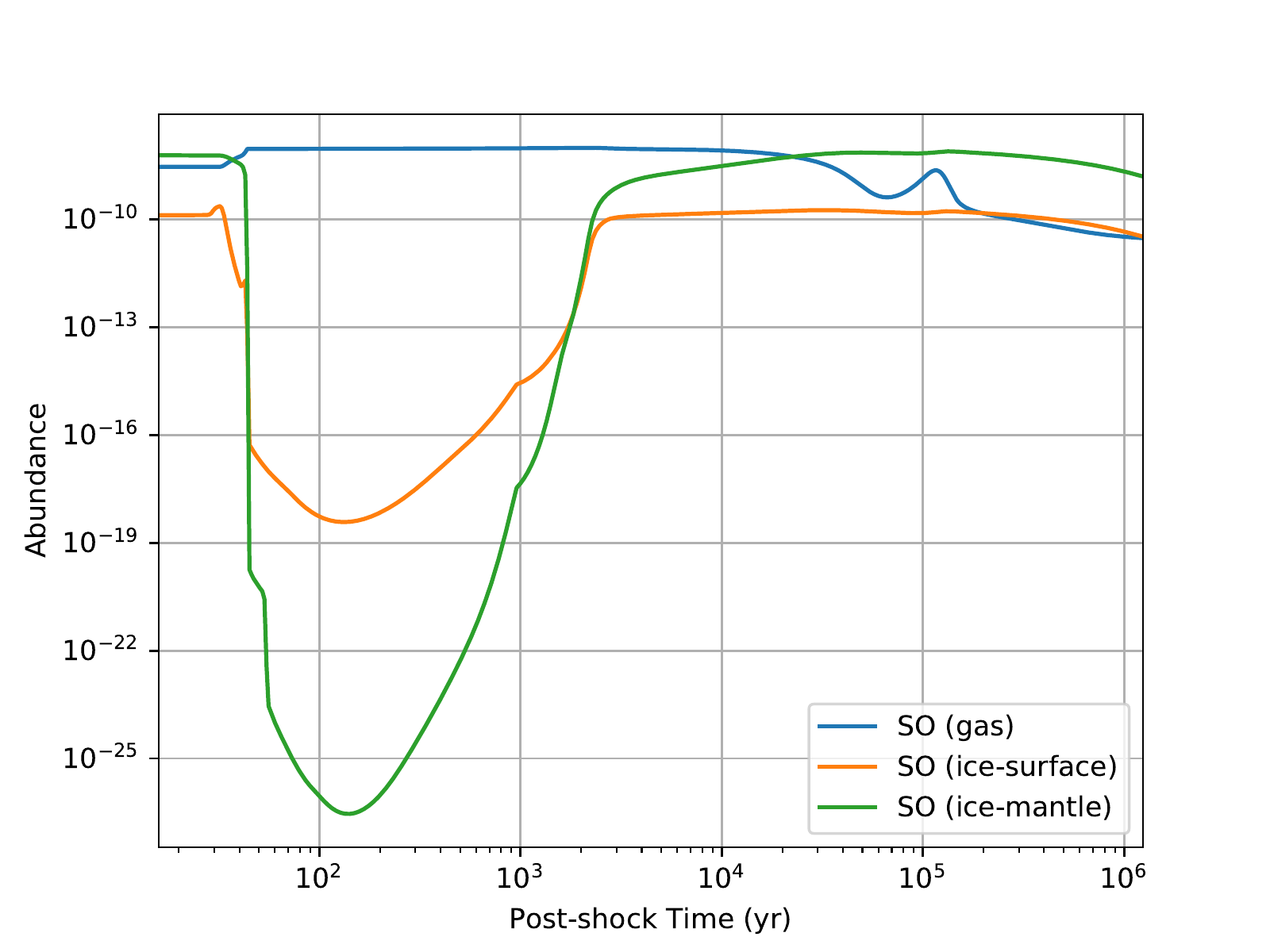}}
\subfigure[\ce{H2S}]{\label{fig:h2s}\includegraphics[width=0.45\textwidth]{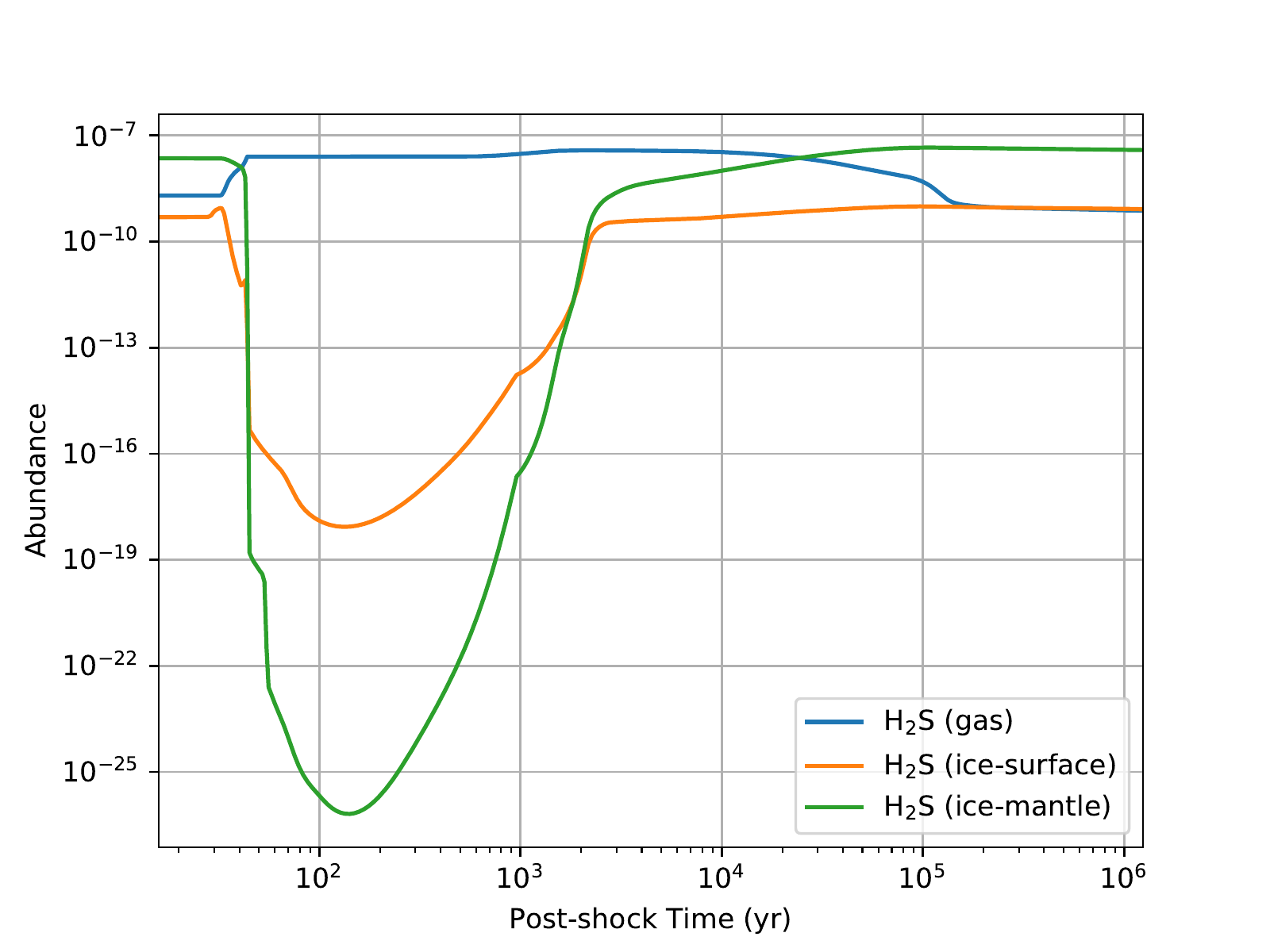}}
\caption{Simulated three-phase abundances of studied species over the time period in which the shock passes through in the model for species that display enhancements during the sputtering, hot gas, and redeposition regimes.} 
\label{fig:3phase8}
\end{figure*}

\begin{figure*}[tb]
\centering     
\subfigure[\ce{SO2}]{\label{fig:so2}\includegraphics[width=0.45\textwidth]{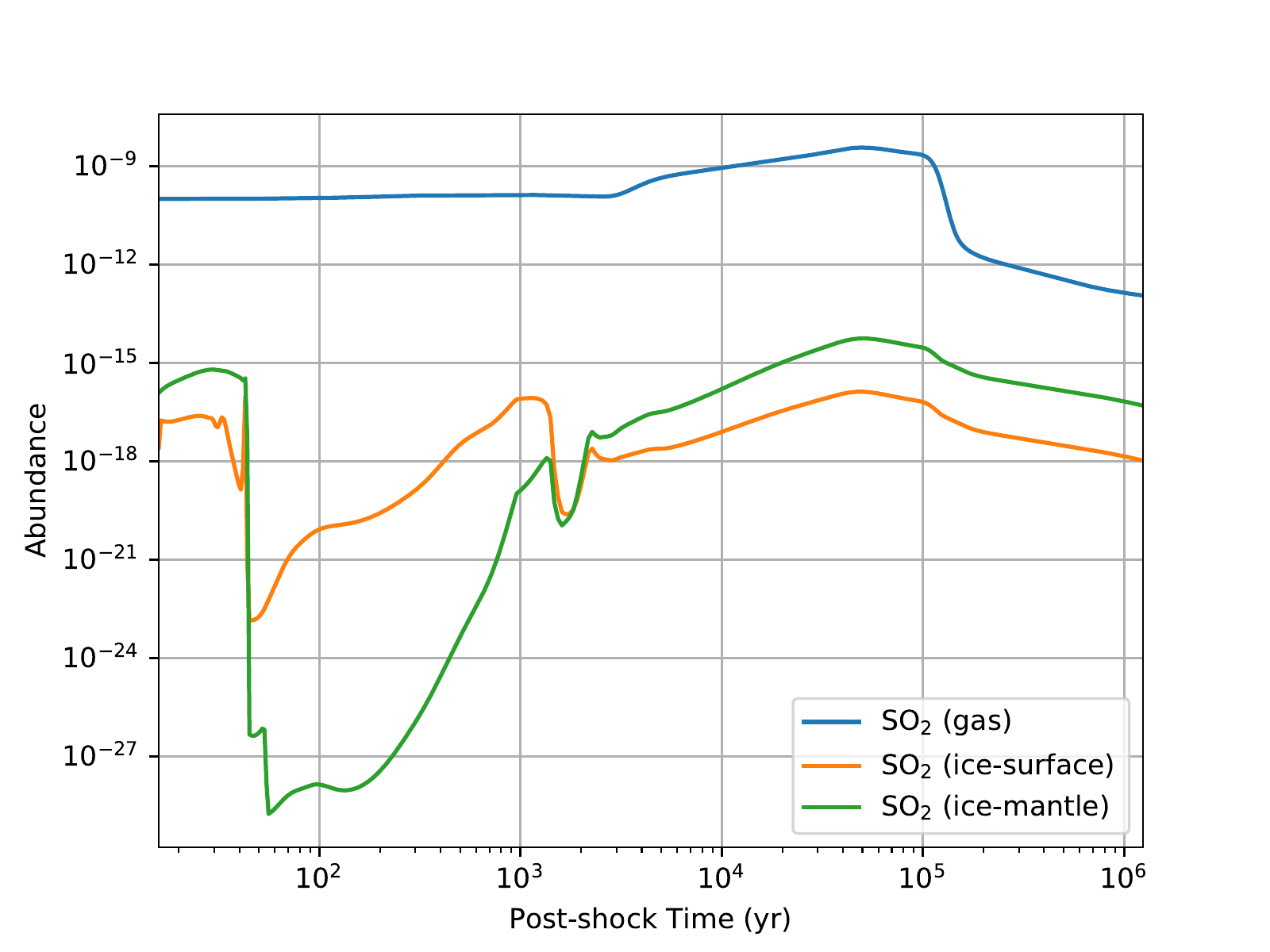}}
\subfigure[\ce{H2CCO}]{\label{fig:h2cco}\includegraphics[width=0.45\textwidth]{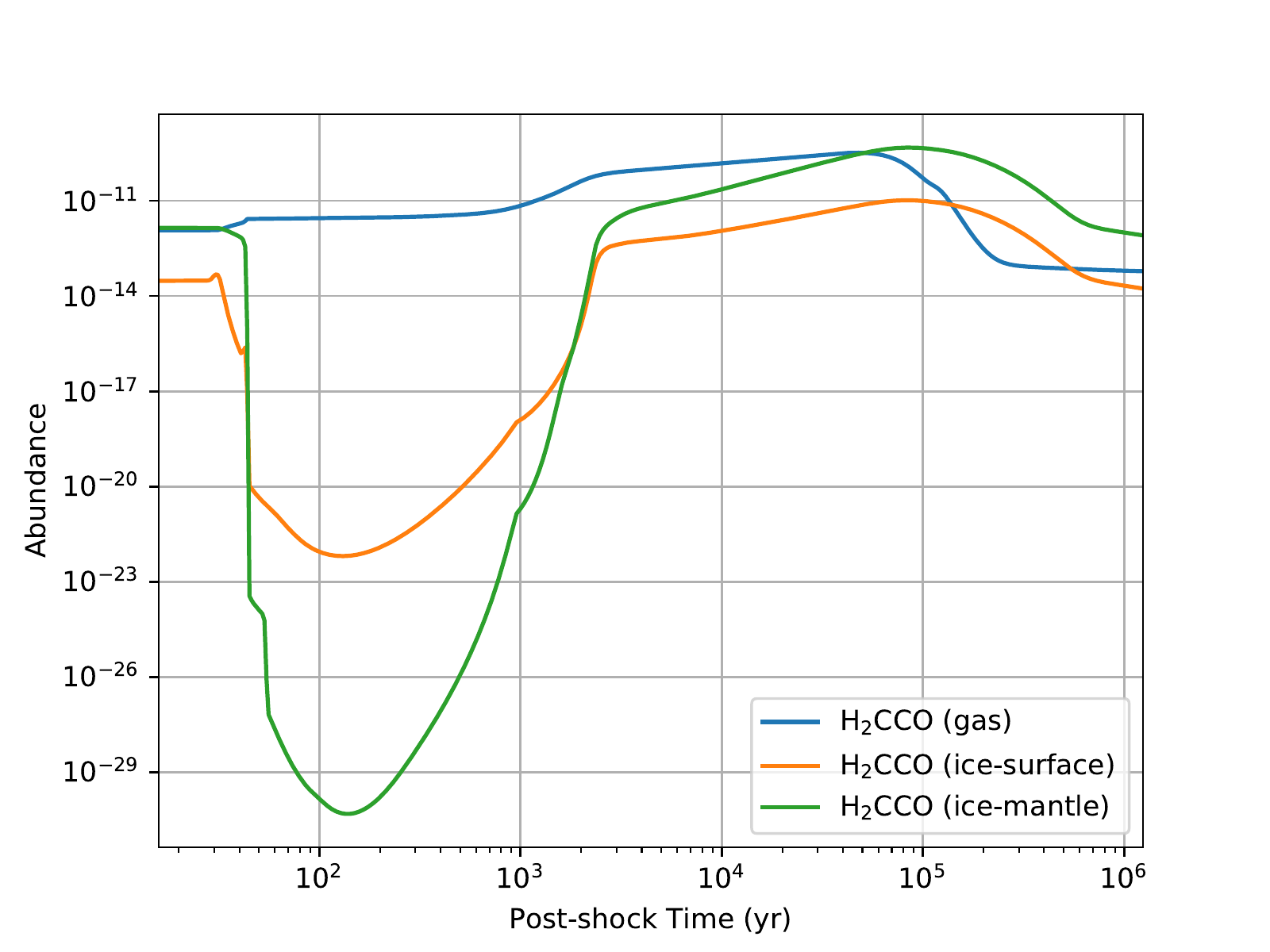}}
\caption{Simulated three-phase abundances of studied species over the time period in which the shock passes through in the model (i.e. $t_{\text{postshock}}$=0 corresponds to $t_{\text{total}}$=10$^6$ years) for species that display strong gas-phase enhancements during the redeposition phase.}
\label{fig:3phase9}
\end{figure*}

\begin{figure*}[htb]
\centering     
\subfigure[\ce{OH}]{\label{fig:oh}\includegraphics[width=0.45\textwidth]{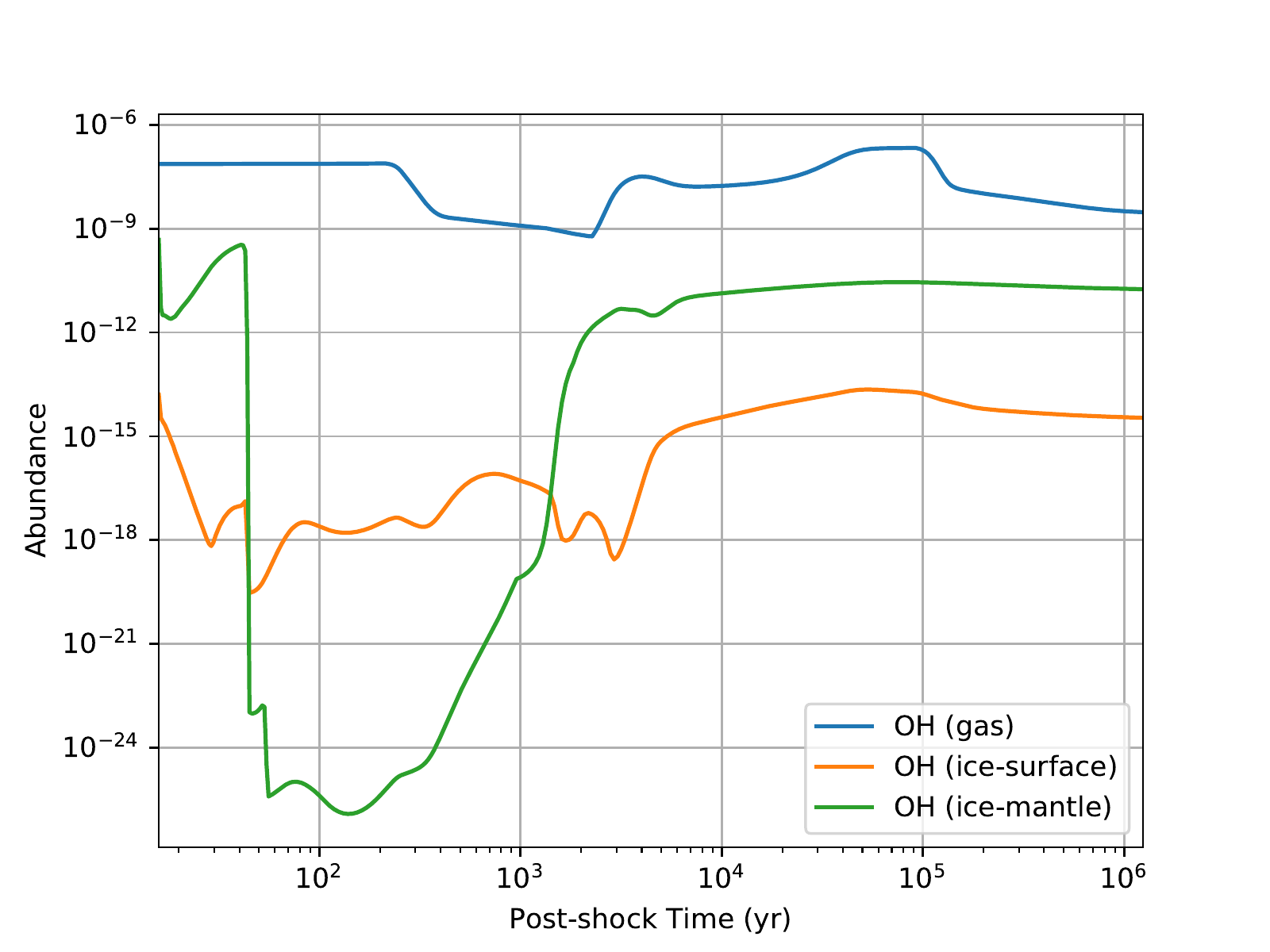}}
\subfigure[\ce{CN}]{\label{fig:cn}\includegraphics[width=0.45\textwidth]{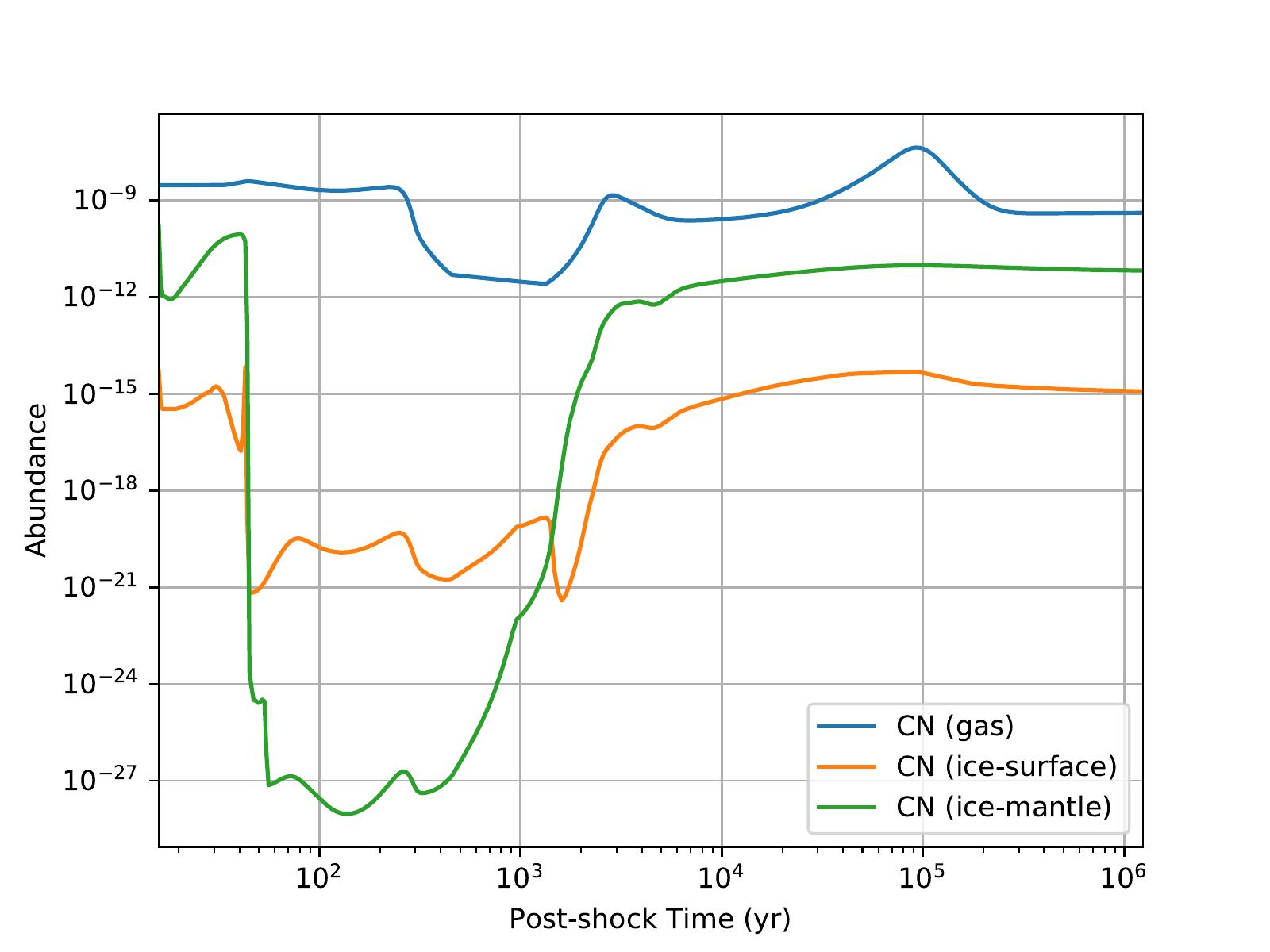}}
\vspace{-1em}\subfigure[\ce{OCN}]{\label{fig:ocn}\includegraphics[width=0.45\textwidth]{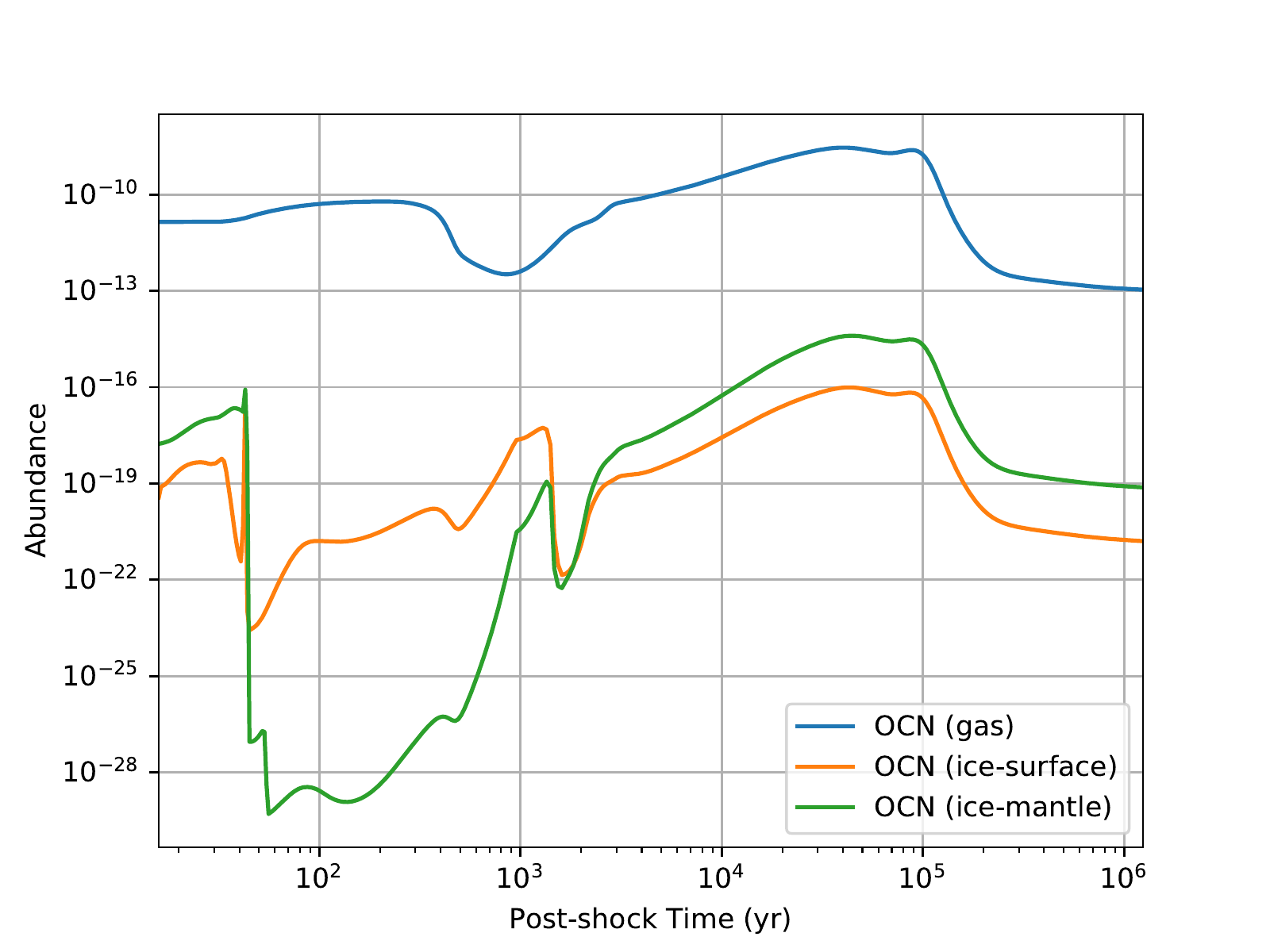}}
\caption{Simulated three-phase abundances of studied species over the time period in which the shock passes through in the model (i.e. $t_{\text{postshock}}$=0 corresponds to $t_{\text{total}}$=10$^6$ years) for species with significant gas-phase depletion during the hot-gas regime.}
\label{fig:3phase10}
\end{figure*}

In this work, we will focus on a collection of molecules that are either classic shock-tracing molecules, key probes of a certain physical process, or molecules that have been detected in L1157 or other similar molecular outflows \citep{Codella:2017ij,Lefloch:2017,Burkhardt:2016bs}. While over 500 species are simulated in the model, this collection will provide sufficient range of formation routes to effectively probe the chemistry within the shocks. 
The three-phase abundances of each of these species over time are shown in Figures \ref{fig:3phase1}-\ref{fig:3phase10}, which have been grouped with a species of similar chemistry, as discussed at the end of this section. Here, we will discuss each regime in detail, discussing collections of molecules that have related profiles and dominate reaction types. 


\subsubsection{Pre-Sputtering Peak Dust Warm Up ($\lesssim 50$ years)}
For the first $\sim$100 years following the shock, the collisions between the gas and grains begin to increase in frequency and impact energy, but the drift velocity has yet to reach its peak value. While the velocities are too low to induce sputtering, it is sufficient to begin heating the dust grain ($\Delta T_{\text{dust}}\lesssim$ 15 K). This can have a number of effects, the foremost being the increased mobility of species to diffuse within the ice. Similar to what is seen in chemical models of hot cores, this warm-up phase can, in turn, rapidly accelerate the formation of complex molecules. This phase is, however, much shorter than the normal timescales for thermal desorption \citep{Garrod:2013id}, which is found to be mostly inefficient in this regime. In this regime, we classify species into one of three types: 
\begin{enumerate}
	\item Those for which gas and ice abundance profiles are constant over time, implying that the dominant ice and gas-phase reactions were already efficient prior to this increase in dust temperature. These species tend to be relatively simple (e.g. CO, H$_2$O, NH$_3$, \ce{CH3OH}), and thus would have relatively low binding energies and could already easily diffuse.  
    \item Those for which the ice abundance rapidly increases in this short time period and the gas-phase abundance becomes enhanced due to non-thermal desorption from exothermic reactions \citep{garrod_non-thermal_2007}. 
These species  (e.g. \ce{CH3CHO}, \ce{OCN}, \ce{CN}, \ce{SO2}) tend to have fairly low ice-abundances compared to what might be expected through observational constraints, which is likely because these species were unable to efficiently diffuse in the dark cloud conditions. The primary formation routes for the molecules were most efficient in the gas-phase, and thus were far too abundant for the ice enhancements to alter it significantly. For some species, like \ce{OH}, the molecule is enhanced on the ice through the dissociation of species, such as HOCO, which also forms CO that reacts with N to form OCN. In the event that the ice-chemistry for these species is found to be more efficient (e.g. comparable or more efficient than the gas-phase chemistry), then it possible that these reactions may induce additional gas-phase enhancements in the peak-sputtering regime. 
    \item Finally, some species also appear to have an enhancement in the ice due to the heating of the dust. However, it only starts occurring once the sputtering is also efficient. This behavior manifests itself as a small increase in the ice-surface abundance prior to the rapid depletion in the next regime. These species tend to be the more complex species in the network, such as \ce{HCOOCH3}, \ce{NH2CHO}, and \ce{HCOOH}, since these require additionally high dust temperatures to effectively diffuse, or were enhanced by the initial production of the ice-species in the previous group.
\end{enumerate}
While a large number of species display some sort of ice enhancement in this short time period, the subsequent chemistry in the more abundant gas phase dominates any additional production during this regime. This implies that in the astrophysical shocks, the secondary chemical effects prior to the peak sputtering are relatively minimal. And to this extent, low-velocity shocks can act as a useful tool into constraining the underlying ice chemistry without having to consider additional physical processes. It should be noted that an expansion of the limited radical-radical reaction network used could impact the results described here.

\subsubsection{Peak-sputtering ($\sim$50-200 years)}
Here, the drift velocity, and thus the sputtering rate reaches its peak value. At this point, the species in the ice surface and mantle are rapidly lifted into the gas phase, leaving only negligible abundances on the grain (i.e. less than 1 complete monolayer). Indeed, this process is seen to be much more efficient that both thermal and reactive desorption within shocks. Again, here, we describe the major groups of species in this regime:
\begin{enumerate}
    \item For highly abundant gas-phase species, such as \ce{OH}, \ce{CS}, and \ce{CN}, the ice abundance was only a small fraction of the total reservoir of those molecules due to gas-phase formation routes being much more efficient. As such, the liberation of the ice population did not appreciably enhance the gas-phase abundance of these molecules. 
	\item For other species, the new gas-phase abundances are nearly equivalent to the pre-shock ice abundance with a near-instantaneous transition. While some of these species may have initially formed in the gas-phase and froze out, the primary reservoir of these species, such as \ce{H2O}, \ce{NH3}, and \ce{HCOOH}, prior to the shock was in these ices. So, the enhancements in the gas-phase abundances are exclusively due to sputtering and not post-shock chemistry. Thus, these species are excellent molecules for constraining their initial ice abundances. 
    \item For many species in the model, such as \ce{CH3OH}, \ce{H2CO}, and \ce{HNCO}, while the post-sputtering gas-phase abundances are equivalent to pre-sputtering ice abundances, the change is not as rapid as for the previous group and the gas-phase enhancement of the molecules begins prior to the majority of the non-thermal desorption brought on by the sputtering. This can be due to either the species efficiently sputtering off before the peak sputtering velocities or additional induced reactions by the increased mobility on the grain and additional enhancements of related species. Observationally, this would mean that the shock-enhanced abundances would correspond to times earlier than when it is seen for the previously mentioned class of molecules. 
    \item Finally, some molecules have gas-phase significant enhancements on top of the sputtering that produce abundances far larger than the pre-sputtering peak solid-phase abundances. Some of these species tend to be the same species that, during the pre-sputtering peak dust heating, would have ice enhancements without the corresponding gas-phase bump (e.g. \ce{OCN} and \ce{CH3CHO}) or were secondary products from the enhancements of other species (e.g. \ce{H2CCO}). Many of these species are enhanced by the initial liberations into the gas-phase of species typically locked in the ice or the slightly-heated gas temperatures. 
\end{enumerate}
The major relic of this phase is clearly the desorption of essentially the entire ice surface and mantle into the gas phase. While some species lift into the gas-phase without additional simultaneous chemical processing, some species are significantly enhanced with the introduction of newly, abundant gas-phase species. From these findings, it is clear that this non-thermal desorption is both highly efficient and produces a unique chemical make up of the gas in the shock relative to other forms of desorption. Furthermore, for a shock velocity of $\sim$20 km s$^{-1}$, this phase would hypothetically occur over roughly at $\sim$500 AU physical scale. And, for nearby molecular outflows such as L1157, $\sim$250 pc away\citep{Looney:2007}, this interior structure within the bow shock would be potentially resolvable with an observation with better than 2$^{\prime\prime}$ resolution, which is easily achievable with modern interferometers. 

\subsubsection{Hot Gas Phase Chemistry ($\sim$200-3$\times$10$^{3}$ years)}
Here, with the ice-surface and mantle now residing in the gas-phase, the gas temperature reaches its peak temperature. With the chemically-enriched gas in a heated environment, new chemistry can proceed that would have been highly inefficient. Due to the extremely low abundances of solid-phase species, the overall profiles of the non-gas-phase abundances will not be discussed here in detail. In general, for a given species the gas-phase abundances will display one of the following trends:
\begin{enumerate}
	\item For some, no appreciable change in the gas-phase abundance occurs in the high temperature regime. Many of the species that were significantly enhanced by the sputtering in the previous regime, such as \ce{CO}, \ce{H2O}, and \ce{NH3}, maintain a stable gas-phase abundance such that there is no significant additional production. This is also true for species that experienced additional enhancements during the peak sputtering, such as \ce{SO} and \ce{SO2}. In particular, \ce{HNCO} does not see any significant post-shock gas-phase enhancements in this regime, which may be due to inefficiencies in the ice-chemistry of the model and will be discussed in a later section.
    \item Other gas-phase abundances are enhanced during the high-temperature regime for species. For some, like \ce{NH2CHO}, \ce{CH3CHO}, and \ce{HCOOCH3}, this is due to an increase of its precursor species due to the sputtering in the previous phase and certain reactions with large barriers or strong temperature-dependencies now being accessible. For others, this enhancement is due to the dissociation of other, more complex species, such as \ce{OH} from the processing of recently enhanced \ce{H2CO}. 
    \item The gas-phase abundances were also diminished for some species, including OH, CN, and OCN. For these, the post-shock chemistry that is enhancing other species is likely depleting these species that were predominantly in the ice prior to this. 
\end{enumerate}
In the shocked outflow L1157, we are able to sample different age shocks by comparing B1 and B2, which differ by about 2$\times$10$^3$ years \citep{Gueth:1996vo}. Thus, the comparison between these two shocks should probe the hot-gas phase of our shock chemistry model. This will be discussed in detail in Section \ref{sec:predictions}.

\subsubsection{Redeposition Phase ($\sim$3$\times$10$^{3}$-10$^{5}$ years)}
Here, the drift velocity and temperatures have fallen back down to the initial pre-shock conditions. Also at this point, the density has finally reached its final value, which is roughly a factor of 5 larger than the pre-shock conditions. Also, the reservoir of molecules on the ice can return to roughly what it was in the initial cold core phase. While the abundances on the grain did increase in the hot-gas regime, it is likely due to the reduction of drift velocity, and thus the repressing of sputtering rather than adsorption. Here, we see:
\begin{enumerate}
	\item Many species have the ice abundances surpass the gas-phase abundances. In addition, over the time-scale in which the ice abundance became larger the gas-phase abundance varies. In many cases, the majority of the gas-phase population is redeposited onto the ice with little additional processing, such as with \ce{CH3OH}, \ce{NH3}, and \ce{H2O}, and occurs very close to 10$^4$ years after the shock. Other species, such as \ce{HNCO} and \ce{NH2CHO}, undergo additional gas-phase enhancements, as their precursors have not yet undergone dissociation or adsorption. As such, the ice-abundances of these species tend to not surpass the gas-phase until closer to 5$\times$10$^4$ years. For many of these species, the ice abundance is either comparable (e.g. \ce{H2O} and \ce{NH3}) or greater than the values prior to the shock (e.g. \ce{NH2CHO}, \ce{CH3CHO}, and \ce{HCOOCH3}).   
    \item Other species, such as \ce{OH}, \ce{OCN} and \ce{SO2}, also have their solid-phase abundances significantly enhanced, but not enough to become close to the gas-phase abundances. Many of these species were predominantly gas-phase species in the pre-shock conditions. And, as such, it is to be expected that the gas-phase reactions would come to dominate again. Meanwhile, some molecules such as \ce{SO} were found to have comparable ice and gas-phase abundances, again due to significant gas-phase production routes in addition to the redeposition.
\end{enumerate}
Overall, essentially every species shows significant rebuilding of the ice mantle. However, the timescale in which species fall back onto the grain was found to vary from species to species, which could be possible to observationally probe for additional post-shock enhancement. While the species with earlier redeposition times and no additional gas-phase production in this regime were found to have comparable ice abundances prior to the peak sputtering, many of the other species had the ice abundances resulting being strongly enhanced (e.g. \ce{HCOOCH3} and \ce{NH2CHO}). As will be discussed in Section \ref{sec:othershockprobes}, this shock-enhancement of the ice may prove to a useful probe for studying the physical and chemical history of these sources.

\subsubsection{Post-shock Dense Gas Phase ($\sim$10\,$^{5}$+ years)}
Beyond 10$^5$ years, the physical conditions do not change over time and represent the effects on the chemistry long after the shock has passed. Excluding the change in density and extinction, the physical conditions are similar to the pre-shock conditions. As such, we should be able to determine whether any changes from pre-shock molecular abundances are due to this change in physical conditions or the after-effects of the shock.
\begin{enumerate}
	\item Many species, at this point, have reached a stable abundance for both the gas phase and ice mantle. This includes the abundant gas-phase species (e.g. \ce{CO} and \ce{HCN}) as well as the main constituents of the ice (e.g. \ce{H2O} and \ce{NH3}). As would be expected, the abundance trends discussed in the previous for these species remain comparable to the pre-shock values.
    \item For many of the most complex species, such as \ce{HCOOCH3} and \ce{CH3CHO}, the ice abundances decrease due to the lack of efficient formation mechanisms on cold grains, as was seen in the pre-shock conditions. This can also been seen by the increase in ice abundance of \ce{CH3OH} and \ce{HCOOH}. Meanwhile, the gas-phase species also decrease due to processing and additional redeposition, but eventually stabilize at an abundance that can be maintained by the gas-phase production routes the species may have. 
    \item Finally, for species such as \ce{OCN} and many of the sulfur-bearing species, both the gas-phase and ice abundances drop like the previous group of molecules. But, because these lighter species are less reliant on the accelerated diffusion on warm grains to form, they eventually stabilize to a final abundance like the first group.
\end{enumerate}
Here, we can see for many of the species that had significant ice-phase enhancements in the first regime of the model, there exists a temporary surge in ice chemistry that lasts for a few times 10$^4$ years, which is longer than many of the gas-phase enhancements commonly seen in these shocks. For a select number of species, their ice abundances are permanently enhanced, which appear to generally be the species that underwent significant post-shock chemistry, like \ce{CH3CHO} and \ce{NH2CHO} or species that were already highly abundant in the ice, but increased in abundance at the end of the model, such as \ce{CH3OH}, \ce{CH3OCH3}, and \ce{HCOOH}. More generally, the entire post-shock regime may be characterized by having ices that have a higher ratio of complex organics relative to \ce{H2O}.  

\subsubsection{Overall Species Classification}\label{sec:classification}
In addition, we found that the evolution of certain clusters of species tended to proceed similarly within the shock, even if they were not directly chemically related, as seen abundance ratios of a single species at different times within the model. First, any overall production or destruction of a molecule during the early onset of the shock can be described through the ratio of the preshock ice abundance, $X_{\text{ice,preshock}}$, and the gas-phase abundance when the gas temperature is at its peak value, $X_{\text{gas,}T_{\text{gas peak}}}$. In the case where sputtering occurs, but no significant post-shock chemistry, the ratio of these abundances, $X_{\text{gas,}T_{\text{gas peak}}}$/$X_{\text{ice,preshock}}$, will be of order unity. When post-shock chemistry creatures further enhancements, such as with \ce{NH2CHO}, this ratio should be large.

\begin{figure}
\centering
\includegraphics[width=\columnwidth]{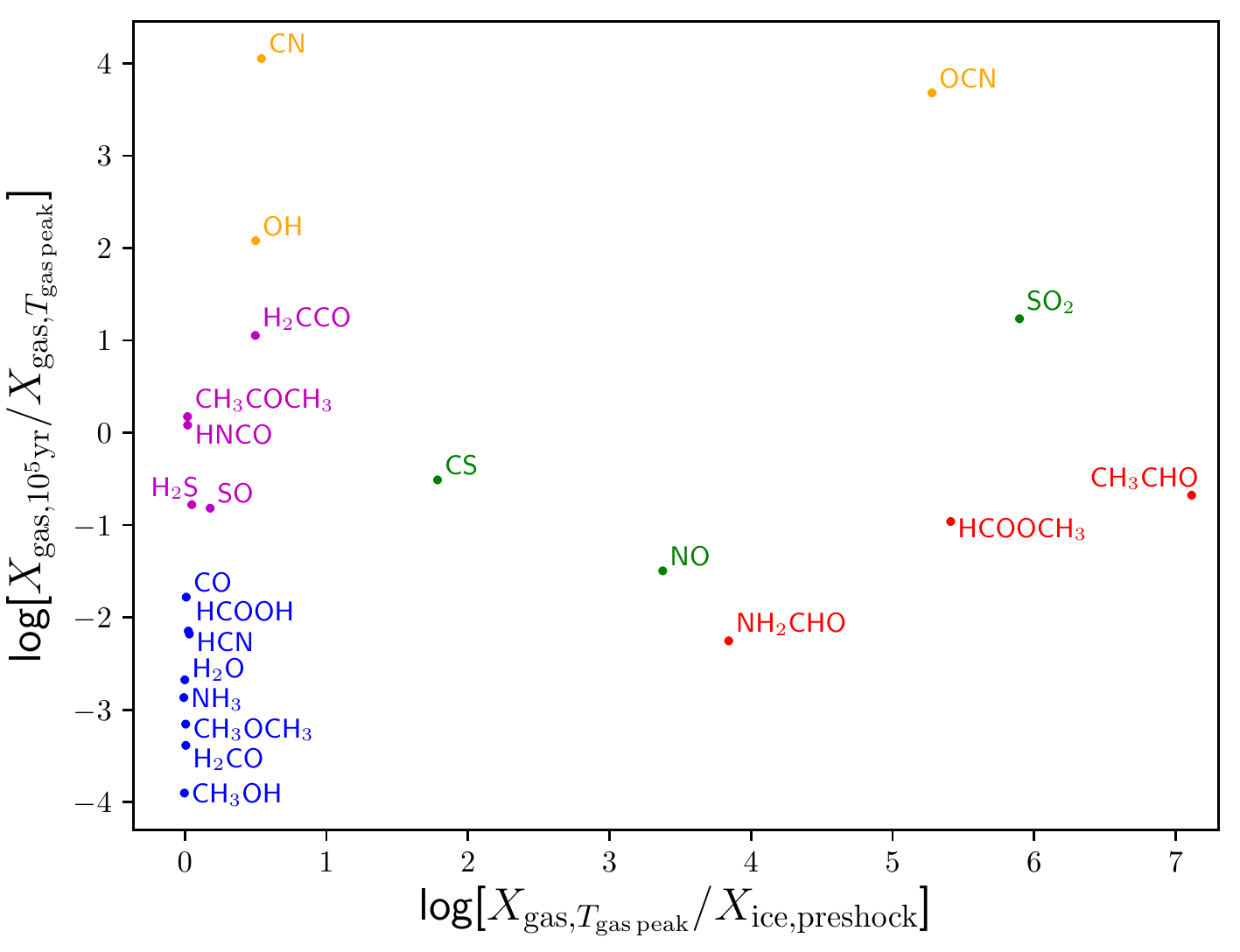} 
\caption{For the species studied, a comparison of the overall enhancement of species within the shock, $X_{\text{gas,}T_{\text{gas peak}}}$/$X_{\text{ice,preshock}}$, and how the gas-phase abundance evolves after the shock, $X_{\text{gas,}10^5\text{yrs}}$/$X_{\text{gas,}T_{\text{gas peak}}}$. Species are colored by classifications given in Section \ref{sec:classification}.}
\label{fig:ratios}
\end{figure}

Second, the timescale in which any gas-phase enhancements or depletion occur can be studied by the ratio between $X_{\text{gas,}T_{\text{gas peak}}}$ and the gas-phase abundance at a time after the initial shock has passed. In this case, we adopt a comparison time $t_{\text{postshock}}=10^5$ years, as it is a key inflection point for many of the species studied here. This ratio, $X_{\text{gas,}10^5\text{yrs}}$/$X_{\text{gas,}T_{\text{gas peak}}}$, should be small (e.g. $<1$) for species where little additional chemistry occurred after sputtered from the grain. For species that still undergo gas-phase chemistry or appear to continue to redeposit well into $10^5$ years after the shock, this ratio can range from order unity to much greater than 1 (e.g. CN). These two ratios stratify the species studied here into general groups defined by their various means of enhancements, as seen in Figure \ref{fig:ratios}.

Many species follow the general bulk evolution of the ice and do not display any significant post-shock gas-phase chemistry, including \ce{CO}, \ce{H2O}, \ce{NH3}, \ce{H2CO}, \ce{CH3OH}, \ce{HCN}, \ce{HCOOH}, and \ce{CH3OCH3}, as seen as blue in Figure \ref{fig:ratios} to have small values for both ratios. Other species, however, display additional enhancements in the gas-phase during the shock events, and many have similar profiles. HNCO, \ce{CH3COCH3}, \ce{H2S}, \ce{H2CCO}, and SO are abundant in the ice prior to the shock, but are both strongly enhanced during the sputtering and redeposition regimes. These species can be seen as a cluster of magenta points in Figure \ref{fig:ratios} to have similarly small values for $X_{\text{gas,}T_{\text{gas peak}}}$/$X_{\text{ice,preshock}}$ but $X_{\text{gas,}10^5\text{yrs}}$/$X_{\text{gas,}T_{\text{gas peak}}}$ are of order unity.  Meanwhile, CS, NO, and \ce{SO2}, shown in green, are primarily gas-phase species that display reasonable enhancements during or just after the redeposition regime. The red-colored species are characterized by strong, consistent post-shock chemistry. \ce{CH3CHO} shows two discrete phases of enhancements, while \ce{HCOOCH3} and \ce{NH2CHO} show strong, continual enhancements from the start of sputtering to redeposition. Both of these groups of species have large values for $X_{\text{gas,}T_{\text{gas peak}}}$/$X_{\text{ice,preshock}}$, but a wider range of $X_{\text{gas,}10^5\text{yrs}}$/$X_{\text{gas,}T_{\text{gas peak}}}$ due to the different timescales of depletion and enhancement. Gas-phase species that tend to be highly reactive, such as \ce{OH}, \ce{CN}, and \ce{OCN}, undergo significant changes overtime, especially characterized by a drop in abundance during the hot gas regime, resulting in large $X_{\text{gas,}10^5\text{yrs}}$/$X_{\text{gas,}T_{\text{gas peak}}}$ values, and are displayed as orange in Figure \ref{fig:ratios}.  As all these species will be enhanced at different discrete stages in the model, these enhancements should be theoretically reproducible by comparing similar-velocity shocks of different ages or the internal structure of single bow shocks.

\section{Discussion} \label{sec:discussion}

\subsection{Time-Sensitive Tracers and Testing Shock-Chemistry Predictions} \label{sec:predictions}
In \citet{Burkhardt:2016bs}, several predictions were made on the nature of the chemistry within L1157 and molecular outflows. Here, we will discuss these predictions, and a few others, in the context of the current results.

\subsubsection{\ce{CH3OH}}\label{sec:ch3oh}
Methanol has long been used as a classical indicator of ``complex'' chemistry, as the six-atom species is thought to be efficiently formed on the ice through successive hydrogenation of the highly abundant CO \citep{herbst_chemistry_2008}. As such, for the purposes of studying the chemistry in shocks, \ce{CH3OH} can be a useful species to study the lifting of the ice into the gas phase. Through deep CARMA observations of L1157, \citet{Burkhardt:2016bs} found that the consistent abundance enhancement throughout the shock event of \ce{CH3OH} indicates that its primary source is from the liberation from grains during the shock. 

In this study, we see that the strongest source of enhancement of gas-phase \ce{CH3OH} is, in fact, the sputtering event with no other significant gas-phase enhancements. On the order of 10$^3$ years after the shock, as probed by the two shock fronts in L1157, the abundance falls off rather smoothly in a relatively log-linear nature between the peak sputtering and the hot gas phase. Thus, on the timescales discussed, it would appear to be a proper tracer to compare the relative shock-enhancements. From the analysis of the full network, other species that may also serve as good tracers of overall shock enhancements include both the major constituent species, such as CO, \ce{H2O}, \ce{NH3}, and additional complex species like \ce{H2CO}, \ce{HCOOH}, and \ce{CH3OCH3}. These species display similar falloff profiles and are all species that should fairly exclusively be produced on the grain in terms of post-shock enhancements. 

While we refrain from discussing species abundances in this model due to the lack of strong constraints on the ice chemistry, it should be noted that the peak phase abundance of \ce{CH3OH} is in agreement with what is typically seen in astrophysical shocks (10$^{-6}$; \citealt{McGuire:2015bp,Burkhardt:2016bs}). However, it has been proposed that dissociation during sputtering may be important and is not considered in this model at this time \citep{Suutarinen:2014}.

In order to quantify this enhancement observationally, the measured abundances will need to be compared another value. While CO is a common species to compare to historically, the majority of the preshock abundance of CO is also locked up in ice. As such, a comparison of CO to other species described here, would probe the rate of desorption and redeposition more than any specific enhancements. As such, a more suitable species to test this would be one that is 1) primarily in the gas phase prior to the shock and 2) whose abundance is either fairly stable or could be described well by a physical model within the source. It may also be possible to compare the \ce{CH3OH} abundance to other positions within the source to determine the degree of enhancement, taking into account the change in physical conditions like density and temperature. For example, if a source contains multiple shock events with similar velocities or a single shock event where \ce{CH3OH} is sufficiently bright where the gas is either not shocked or the shock has long passed, the overall abundance profile of the molecule can be compared to modeled profiles. 

\subsubsection{HNCO}
Due to the abundance enhancements of HNCO in the L1157 bow shocks, especially comparing the B1 shock front to the older B2 shock front ($\Delta t\sim$2000 years), it has been proposed that HNCO may be enhanced by post-shock gas-phase chemistry in addition to the initial release from sputtering \citep{Burkhardt:2016bs,RodriguezFernandez:2010dp}. Specifically, HNCO was seen to have a factor of two enhancement in the abundance over this time scale. In the model, however, at 2$\times$10$^{3}$ years after the shock (i.e. during the hot-gas regime), by comparing the abundances between the beginning of this phase ($\sim$200 years) and at the peak ($\sim$2.1$\times$10$^3$ years), one can see that this enhancement is only $\sim$20\%. While we do not reproduce this peak, it is likely because the dominant reactions to produce HNCO in this phase require CH$_2$ or OCN, both of which were depleted prior to the peak of HNCO. In addition, OCN seems to be underproduced on the ice relative to estimated abundances from infrared absorption ice studies \citep{Boogert:2015fx}. It should be noted that the calculated ice abundance of OCN is comparable to observed gas-phase abundances in dense cores \citep{Marcelino:2018}. As such, while we do not reproduce this post-shock enhancement, it is possible that is may still be feasible if the pre-shock ice abundance of OCN was closer to observed values.

Despite this lack of overall abundance enhancement hot-gas chemistry regime, the chemistry of HNCO is unique compared to the sputtering-dominated species, as discussed in Section \ref{sec:classification} and seen in Figure \ref{fig:ratios}. Instead of a strong enhancement, as seen by \ce{NH2CHO}, the gas-phase abundance of HNCO merely remains elevated relative to the preshock value for much longer than the sputtering-dominated species, such as \ce{CH3OH}. As a result, the HNCO/\ce{CH3OH} abundance ratio increases as the shock gets older until HNCO fully redeposits onto the ice, as was described by \citep{Burkhardt:2016bs}.

\subsubsection{Sulfur Chemistry}
It has been proposed that the chemistry of sulfur-bearing molecules may be a useful indicator of shock evolution \citep{Hartquist:1980, Neufeld:1989, Codella:2005, Wakelam:2005}, including an increase of the \ce{SO2}/\ce{SO} ratio over time and the general conversion of \ce{H2S} into \ce{SO} and \ce{SO2}. Here, while both \ce{SO} and \ce{SO2} are enhanced at timescales $>$10$^{3}$ years, we find that \ce{SO2} increases significantly more rapidly over this time, providing further evidence to support this earlier hypothesis. However, \ce{H2S} does also appear to be enhanced in the later stages of the shock, which suggests that it is not a primary source of sulfur for \ce{SO} and \ce{SO2}. As it is apparent that sulfur chemistry can be strongly tied with the evolution of a shock, a more in depth study of high temperature sulfur chemistry may be warranted.


\subsection{Other useful shock-probes}\label{sec:othershockprobes}
Through observations of the internal structure of the B1 shock front, \citet{Codella:2015kq,Codella:2017ij} probed the shock chemistry of \ce{HDCO}, \ce{CH3CHO}, and \ce{NH2CHO} by studying the morphology of the enhanced abundances within a single bow shock, suggesting that the abundance of shock tracers may be segmented by different phases of chemistry. While \ce{HDCO} was found to be primarily enhanced at the very head of the bow shock, \ce{NH2CHO} and \ce{CH3CHO} peaked at a location slightly behind the shock front, suggesting that their shock enhancements were temporally delayed relative to the initial shock event (i.e. post-shock, gas-phase chemistry). Here, we see that the distribution of \ce{CH3CHO}, \ce{NH2CHO}, and additionally \ce{HCOOCH3}, are continuously enhanced in the shock through gas-phase reactions. \ce{NH2CHO} and \ce{HCOOCH3}, as well, show enhancements due to the sputtering event, but additional gas-phase processes are clearly efficient simultaneously to the sputtering around 100 years after the shock begins. It would therefore be a useful test to see if \ce{HCOOCH3}, or other molecules of similar complexity, display any similar distributions to that of \ce{NH2CHO} and \ce{CH3CHO}.


Furthermore, the shock-induced enhancement of the ice abundances, both permanent (e.g. \ce{CH3OH}, \ce{H2CO}, and \ce{HCOOH}) and temporary (e.g. \ce{NH2CHO} and \ce{HNCO}) could prove to be very useful to provide evidence for prior shock events in a given source. With the upcoming launch of the James Webb Space Telescope \citep{Kirkpatrick:2017}, infrared absorption observations of ice features may be capable of differentiating between the shock-enhanced ices and ices with a relatively quiescent history. Because these ice enhancements remain far longer than the gas-phase enhancements and after the shock has passed, this effect could also provide evidence for the frequency in which shocks occur in the interstellar medium and whether a star forming region was triggered by shocks or general gravitational collapse. 

\subsection{Comparison of shock velocities}
Here, we focused on a shock velocity of 20 km s$^{-1}$. However, additional models were run over the range of shock velocities typical for C-shocks (10-40 km s$^{-1}$) in order to confirm that the conclusions still hold in a wider range of parameter space. While small differences do arise due to changes in the shock timescales, efficiencies of grain heating, and the peak temperature of the shocked gas, the ultimate time-dependent profiles remain mostly consistent to the extent that there are no major disagreements with the conclusions discussed here. A more in-depth study of how the chemistry changes across shock velocities will follow in future work.

\subsection{Application for Protoplanetary Disks}
In protoplanetary disks, the drift velocities observed in shocks tend to be lower than in molecular outflows \citep{Ilee:2017}. Thus, sputtering may not be as efficient as a source for non-thermally desorbing grain species in these regions. However, as we showed, additional ice chemistry can be induced through shocks by the heating of the dust grains, thus increasing thermal diffusive chemistry, with some fraction of the products of these exothermic reactions undergoing reactive desorption.


\subsection{Future Directions}
In order to fully constrain the complex interplay between the shock and the resulting chemistry, future work will focus on exploring the parameter space of the model; testing to see whether species are still able to efficiently sputter at very low velocity shocks or how the initial conditions of the dark cloud  (e.g. elemental abundances, cloud age, ice composition) affects these results. 

Many of the shock-induced enhancements can be observationally tested through interferometric or infrared ice absorption techniques. As such, dedicated studies to gather a rigorous sample of shock conditions and ages would allow this model to be fully constrained and improved.

In addition, more robust treatments of several processes could be relatively easily implemented to expand the power of the model's predictions. This includes the refractory grain as a fourth chemical phase to be eroded during high-velocity shocks, which would allow us to also consider the effect shock chemistry has on the prototypical shock tracer, SiO \citep{Gusdorf:2008js}. Also, we can include a more robust treatment of the dust size distribution, which has been shown to affect the chemistry in young stellar objects \citep{harada:2017}, and how that is impacted by the shock and the erosion of the ice mantle. The sputtering yields for many of these species have yet to be definitely and robustly studied \citep{Dartois:2018}. And, additional experimental and robust molecular dynamics calculations could allow for significant non-thermal desorption to be used as a probe of unconstrained ice chemistry \citep{Cassone:2018}. Furthermore, additional high temperature chemistry can be included, including chemisorption on bare grains and additional hot gas-phase reactions with barriers, to better study the effects of the post-shock gas-phase chemistry. Similarly, in addition to sputtering, the collisions with high-velocity particles could also cause molecules to dissociate \citep{Nesterenok:2018}, which could increase the volatility of the post-shock gas. 

Reactions in dust grain ice-mantles driven by cosmic ray bombardment have also been found to be highly efficient at enhancing the chemistry in dark clouds \citep{shingledecker_cosmic_2018}. The inclusion of these processes may more accurately reproduce observed ice abundances in our pre-shock ice build up. Moreover, the enhanced flux of cosmic rays within astrophysical shocks will likely also impact the abundances of many species both by enhancing desorption via processes such as sputering while simultaneously driving additional chemistry through collisions with the gas and dust grains. Thus, testing the extent to which the cosmic ray ionization rate can impact the pre- and post-shock chemistry will also be a promising avenue of study in the future. Similarly, interactions with cosmic rays are known to affect both dust grain charging and heating, therefore warranting a more robust treatment of the grain charge would also likely be important \citep{ivlev_interstellar_2015}.

\section{Conclusion}
Through inclusion of sputtering, shock-physics parameters, high temperature reactions, and shock-induced dust heating into the augmented three-phase gas-grain chemical network code \textsc{nautilus}, we are able to reproduce many of the predictions made by \citet{Burkhardt:2016bs}. In general, we find that some species will uniquely trace the overall evolution of the ice and have minimal post-shock gas-phase chemistry. Others display significant enhancements between the regimes of peak sputtering, hot gas-phase, and the redeposition onto the ice. And still other species already exist in the gas, but interactions with the shocks can impact their temporal abundance evolution. From this, we discussed how methanol, among other studied species, can serve to be a useful probe of the underlying ice chemistry. In addition, the temporary gas-phase enhancements can be easily tested observationally by modern observatories at the necessary angular resolution to correspond to the timescales discussed here. Finally, we find that the ice abundance can become permanently altered long after the shock, which may be possible to observe with the James Webb Space Telescope.




\section{Acknowledgments}

A.M.B. was a Grote Reber Fellow for a portion of this project, and support for this work was provided by the NSF through the Grote Reber Fellowship Program administered by Associated Universities, Inc./National Radio Astronomy Observatory and the Virginia Space Grant Consortium.
A.M.B. also acknowledges support from the Smithsonian Institution as a current Submillimeter Array (SMA) Fellow. 
A.M.B. would like to thank Karin \"{O}berg for helpful discussions on ice chemistry.
C.N.S. wishes to thank the Alexander von Humboldt Foundation for support. 
Support for B.A.M. was provided by NASA through Hubble Fellowship grant \#HST-HF2-51396 awarded by the Space Telescope Science Institute, which is operated by the Association of Universities for Research in Astronomy, Inc., for NASA, under contract NAS5-26555. 
E. H. thanks the National Science Foundation for support of his astrochemistry program.



\end{document}